\documentclass[12pt]{article}

\usepackage{epsfig}
\usepackage{rotating}
\usepackage{a4p}

\newcommand{\etal}      {{\it et~al.}}
\newcommand{\PhysLett}  {Phys.~Lett.}

\newcommand{\PhysRev}   {Phys.~Rev.}
\newcommand{\PhysRevl}  {Phys.~Rev.~Lett.}
\newcommand{\NPhys}     {Nucl.~Phys.}
\newcommand{\ZPhys}     {Z.~Phys.}

\newcommand{\ee}        {\mbox{$\rm e^+e^-\ $}}

\newcommand{\pp}        {\rm p\={p}}

\newcommand{\Qto}       {Q_{\mathrm {t_{out}}}}
\newcommand{\Qts}       {Q_{\mathrm {t_{side}}}}
\newcommand{\Ql}        {Q_{\mathrm \ell}}
\newcommand{\qz}        {q_0}
\newcommand{\qt}        {q_{\mathrm t}}
\newcommand{\ql}        {q_{\mathrm \ell}}
\newcommand{\kt}        {k_{\mathrm t}}
\newcommand{\Rto}       {R_{\mathrm {t_{out}}}}
\newcommand{\Rts}       {R_{\mathrm {t_{side}}}}
\newcommand{\Rlong}     {R_{\mathrm {long}}}
\newcommand{\Rlongto}   {R_{\mathrm {long,t_{out}}}}
\newcommand{\Rz}        {R_0}
\newcommand{\Rt}        {R_{\mathrm t}}
\newcommand{\Rl}        {R_{\mathrm \ell}}
\newcommand{\eto}       {\epsilon_{\mathrm {t_{out}}}}
\newcommand{\ets}       {\epsilon_{\mathrm {t_{side}}}}
\newcommand{\elong}     {\epsilon_{\mathrm {long}}}
\newcommand{\dz}        {\delta_0}
\newcommand{\dt}        {\delta_{\mathrm t}}
\newcommand{\dl}        {\delta_{\mathrm \ell}}
\newcommand{\YYK}       {Y_{\mathrm {YK}}}

\begin{document}
\begin{titlepage}
\begin{center}
{\Large EUROPEAN ORGANIZATION FOR NUCLEAR RESEARCH}
\end{center}
\bigskip\bigskip
\begin{flushright}
{\Large CERN-PH-EP/2007-025}\\
{\Large OPAL PR423}\\
{\Large 9 July 2007}
\end{flushright}
\bigskip
\begin{center}{\Huge\bf
Bose-Einstein study of\\
position-momentum correlations\\
of charged pions\\
in hadronic Z$^0$ decays}
\end{center}
\bigskip\medskip
\begin{center}
{\huge The OPAL Collaboration}
\end{center}
\bigskip\bigskip\bigskip
%
%
\begin{center}{\Large\bf Abstract}\end{center}
%
%
A study of Bose-Einstein correlations in pairs of identically charged
pions produced in e$^+$e$^-$
annihilations at the Z$^0$ peak has been performed for the
first time assuming a non-static emitting source.
The results are based on the high statistics data obtained with the
OPAL detector at LEP.
The correlation functions have been analyzed in intervals of the average
pair transverse momentum and of the pair rapidity, in order to study
possible correlations between the pion production points and their momenta
(position-momentum correlations).
The Yano-Koonin and the Bertsch-Pratt parameterizations have been fitted to
the measured correlation functions to estimate the geometrical parameters
of the source as well as the velocity of the source elements with respect
to the overall centre-of-mass frame.
The source rapidity is found to scale approximately with the pair
rapidity, and both the longitudinal and transverse source dimensions
are found to decrease for increasing average pair transverse momenta.
\bigskip\bigskip\bigskip\bigskip
\begin{center}
{\Large (Submitted to Eur. Phys. J. C)}
\end{center}
\end{titlepage}
\newpage
\begin{center}
{\Large The OPAL Collaboration}
\end{center}
\bigskip
\begin{center}{
G.\thinspace Abbiendi$^{  2}$,
C.\thinspace Ainsley$^{  5}$,
P.F.\thinspace {\AA}kesson$^{  7}$,
G.\thinspace Alexander$^{ 21}$,
G.\thinspace Anagnostou$^{  1}$,
K.J.\thinspace Anderson$^{  8}$,
S.\thinspace Asai$^{ 22}$,
D.\thinspace Axen$^{ 26}$,
I.\thinspace Bailey$^{ 25}$,
E.\thinspace Barberio$^{  7,   p}$
% ...
}\end{center}
\bigskip\bigskip
$^{  1}$School of Physics and Astronomy, University of Birmingham,
Birmingham B15 2TT, UK
\newline
$^{  2}$Dipartimento di Fisica dell' Universit\`a di Bologna and INFN,
I-40126 Bologna, Italy
\newline
% ...
$^{  5}$Cavendish Laboratory, Cambridge CB3 0HE, UK
\newline
% ...
$^{  7}$CERN, European Organisation for Nuclear Research,
CH-1211 Geneva 23, Switzerland
\newline
$^{  8}$Enrico Fermi Institute and Department of Physics,
University of Chicago, Chicago IL 60637, USA
\newline
% ...
$^{ 21}$Department of Physics and Astronomy, Tel Aviv University,
Tel Aviv 69978, Israel
\newline
$^{ 22}$International Centre for Elementary Particle Physics and
Department of Physics, University of Tokyo, Tokyo 113-0033, and
Kobe University, Kobe 657-8501, Japan
\newline
% ...
$^{ 25}$University of Victoria, Department of Physics, P O Box 3055,
Victoria BC V8W 3P6, Canada
\newline
$^{ 26}$University of British Columbia, Department of Physics,
Vancouver BC V6T 1Z1, Canada
\newline
% ...
$^{  p}$ now at The University of Melbourne, Victoria, Australia
\newline
% ...
\newpage
%
%
\section{Introduction}
%
%
The space-time evolution of a source emitting particles can be probed
using intensity interferometry.
Bose-Einstein correlations (BECs) in pairs of identical bosons have been
studied at different centre-of-mass energies and for different initial
states (\ee~\cite{epem}, pp and \pp~\cite{pantip}, lepton-hadron~\cite{lephad},
nucleus-nucleus collisions~\cite{h_ions}).
BECs manifest themselves as enhancements in the production of identical
bosons which are close to one another in phase space.
They can be analysed in terms of the correlation function
\begin{equation}
C(p_1,p_2) = \frac{\rho(p_1,p_2)}{\rho_0(p_1,p_2)}
\end{equation}
where $p_1$ and $p_2$ are the 4-momenta of the two bosons,
$\rho(p_1,p_2)$ is the density of the two identical bosons and
$\rho_0(p_1,p_2)$ is the two-particle density in the absence of
BECs (reference sample).
From the experimental correlation function one can extract the dimension of
the source element (frequently called correlation length or radius of the
emitting source), i.e. the length
of the region of homogeneity
from which pions are emitted that have momenta similar enough to
interfere and contribute to the correlation function.
\par
At LEP Bose-Einstein correlations were analysed extensively in Z$^0$
hadronic events~[5-11].
Two-pion correlations were studied as a function of the relative
4-momentum $q=(p_1-p_2)$ of the pair:
$C(p_{1},p_{2})$ = $C(q)$.
It was found that the radius of the emitting region, supposed
spherical, is of the order of 1 fm and increases with the number of
jets in the event~\cite{fierro}.
No significant differences were observed in the source dimensions between
the $\pi^{\pm}\pi^{\pm}$ and the $\pi^{0}\pi^{0}$ systems~\cite{pi0}; on
the other hand, smaller radii were measured in K$^{\pm}$K$^{\pm}$
and K$^{0}$K$^{0}$
($\overline{\mathrm K^0}\overline{{\mathrm K^0}}$) pairs
compared with pion pairs~\cite{mass_dep}.
Genuine three-pion BECs were also observed~\cite{trepi}.
Up to fifth-order genuine correlations of identically charged pions
were obtained by OPAL~\cite{sarki}, where BECs were shown to be an essential
ingredient of the correlation scaling observed there, also for
all-charged higher-order correlations.
The hypothesis that the source is spherical was tested studying the
correlations in terms of components of $q$: two- and three-dimensional
analyses have shown that the pion emission region is elongated rather than
spherical, with the longitudinal dimension, along the event thrust axis,
larger than the transverse one~\cite{tr_long,tr_long_al}.
BECs were also studied in \ee$\rightarrow~$W$^+$W$^-$ events: no evidence
of correlations between pions originating from different W bosons was
found~\cite{ww}.
\par
All the results listed above were obtained under the hypothesis that the
momentum distribution of the emitted particles is homogeneous throughout
the source elements, as would happen if the source is static.
In the case of a dynamic, i.e. expanding, source, the dimension of the
regions of homogeneity varies with the momentum of the emitted particles.
The expansion leads to correlations between the space-time
emission points and the particle 4-momenta (position-momentum correlations)
which generate a dependence of the BEC radii on the pair momenta.
In this case, the correlation function is expected to depend on the average
4-momentum of the pair $K=(p_1+p_2)/2$ in addition to the relative
4-momentum $q$: $C(p_1,p_2)=C(q,K)$~\cite{time_rel},
so that the measured radii correspond to regions of homogeneity
in $K$, i.e. effective source elements of pairs with momentum $K$.
\par
Published investigations of the source dynamics in \ee collisions are
available at energies lower than LEP's~\cite{epem}.
A dependence of the source radii on different components of the 4-vector
$K$ has been observed in more complex systems such as the emission
region created after a high-energy collision between heavy nuclei.
In particular, source radii have been found to decrease for increasing
pair transverse momenta $\kt$ (or, equivalently, transverse masses
$m_{\mathrm t} = \sqrt{ k_{\mathrm t}^2 + m_{\pi}^2 }$)~\cite{h_ions}.
Hydrodynamical models for heavy ion collisions~\cite{hydro} explain this
correlation in terms of an expansion of
the source, due to collective flows generated by pressure gradients.
A similar dependence of the size parameters on $m_{\mathrm t}$ was measured in
pp collisions~\cite{star_pp}.
Longitudinal position-momentum correlations can be expected in \ee
annihilations as a consequence of string fragmentation~\cite{string}.
Models based on different assumptions (the Heisenberg
uncertainty principle~\cite{alexander}, the generalized Bjorken-Gottfried
hypothesis~\cite{bialas}) predict radii decreasing with the transverse
mass also for sources created in \ee collisions.
\par
In this paper, which continues a series of OPAL studies on
BECs~\cite{fierro,tr_long}, a measurement of three-dimensional
Bose-Einstein correlation functions is presented and the correlation
functions are analyzed in order to measure their dependence on $K$ and
investigate potential dynamical features of the pion-emitting source created
after an \ee annihilation at a centre-of-mass energy of about 91 GeV.
%
%
\section{Experimental procedure}
%
%
A detailed description of the OPAL detector can be found
in~\cite{opaldet,opaldet1}.
In the present analysis, we have used the same data sample, about 4.3
million multihadronic events from Z$^0$ decays, and have applied the
following selection cuts on tracks and events, identical to the ones
described in~\cite{tr_long}.
First, the event thrust axis was computed, using tracks with a
minimum of 20 hits in the jet chamber, a minimum transverse momentum of 
150 MeV and a maximum momentum of 65 GeV.
Clusters in the electromagnetic calorimeter are used if their energies
exceed 100 MeV in the barrel or 200 MeV in the endcaps.
Only events well contained in the detector were accepted,
requiring $|{\rm cos}\theta_{\mathrm {thrust}}|<0.9$,
where $\theta_{\mathrm {thrust}}$ is
the polar angle of the thrust axis with respect to the beam axis
\footnote{The coordinate system is defined so that $z$ is the
coordinate parallel to the e$^+$ and e$^-$ beams, with
positive direction along the e$^-$ beam; $r$ is the
coordinate normal to the beam axis, $\phi$ is the azimuthal angle and
$\theta$ is the polar angle with respect to +$z$.}.
Then, a set of cuts, specific to BEC analyses, were applied.
Tracks were required to have a maximum momentum of 40 GeV and to
originate from the interaction vertex.
Electron-positron pairs from photon conversions were rejected.
Events were selected if they contained a minimum number of five 
tracks and if they were reasonably balanced in charge, i.e. requiring 
$|n_{\mathrm {ch}}^{+}-n_{\mathrm {ch}}^{-}|/(n_{\mathrm {ch}}^{+}+
n_{\mathrm {ch}}^{-}) \leq 0.4$,
where $n_{\mathrm {ch}}^{+}$ and $n_{\mathrm {ch}}^{-}$ are the number
of positive and negative charge tracks, respectively.
About 3.7 million events were left after all quality cuts.
All charged particle tracks that passed the selections were used,
the pion purity being approximately 90\%.
No corrections were applied for final state Coulomb interactions.
All data and Monte Carlo distributions presented here are given at the
detector level, i.e. not corrected for effects of detector acceptance
and resolution.
\par
The correlations were measured as functions of two different sets
of variables, components of the pair 4-momentum difference $q$ in two
different frames.
\par The first set, ($\Ql,\Qts,\Qto$),
was evaluated in the Longitudinally CoMoving System (LCMS)~\cite{lcms_c}.
For each pion pair, the LCMS is the frame, moving along the thrust axis, in
which the sum of the two particle momenta,
$\vec{p}$=($\vec{p}_1$+$\vec{p}_2$),
lies in the plane perpendicular to the event thrust axis.
The momentum difference of the pair, $\vec{Q}=(\vec{p}_1-\vec{p}_2)$
is resolved into the moduli of the transverse component,
$\vec{Q}_{\mathrm t}$, and of the longitudinal component,
$\vec{Q}_{\mathrm \ell}$, where the longitudinal
($\hat{\ell}$) direction coincides with the thrust axis.
$\vec{Q}_{\mathrm t}$ may in turn be resolved into ``out", $\Qto$,
and ``side", $\Qts$, components
\begin{equation}
\vec{Q}_{\mathrm t} =
\Qto \hat{o} + \Qts \hat{s}
\end{equation}
where $\hat{o}$ and $\hat{s}$ are unit vectors in the plane perpendicular
to the thrust direction, such that 
$\vec{p} = p \hat{o}$ defines the
``out" direction and $\hat{s}=\hat{\ell}\times\hat{o}$ defines the 
``side" direction.
It can be shown~\cite{cross_term} that, in the LCMS, 
the components $\Qts$ and $\Ql$ reflect only the difference 
in emission space of the two pions, while $\Qto$ depends on the 
difference in emission time as well.
\par The second set,
$(\qt,\ql,\qz)$, was evaluated in the event
centre-of-mass (CMS) frame.
For each event, two hemispheres are defined by the plane perpendicular
to the thrust axis.
Each pair is then associated to the hemisphere containing the vector sum
of the three-momenta.
The pair 4-momentum difference $q$ is resolved into the energy 
difference $\qz = (E_1 - E_2)$ and the 3-momentum difference
$\vec{q}=(\vec{p}_1-\vec{p}_2)$.
The vector $\vec{q}$ is further decomposed into $\qt$ and
$\ql$, the transverse and longitudinal components, respectively, with
respect to the thrust axis.
In each pair, index 1 corresponds to the particle with the highest energy,
so that $\qz \geq 0$.
The longitudinal component, $\ql$, may be either positive, in case the
vector difference $\vec{q}$ lies in the pair hemisphere, or
negative, in the opposite case.
The transverse component, $\qt$, is positive definite.
\par The experimental three-dimensional correlation functions $C$ are
defined, in a
small phase space volume around each triplet of
$\Ql$, $\Qts$ and $\Qto$
(or $\qt$, $\ql$ and $\qz$)
values, as the number of like-charge pairs in that volume divided by the
number of unlike-charge pairs:
\begin{equation}
C=
\frac{N_{\pi^+ \pi^+} + N_{\pi^- \pi^-}}{N_{\pi^+ \pi^-}} =
\frac{N_{\rm like}}{N_{\rm unlike}}.
\end{equation}
In order to have adequate statistics in each bin, a bin size of 40 MeV was
chosen in each component of $q$, which is larger than the estimated
detector resolution of 25 MeV~\cite{fierro}.
\par
Long-range correlations are present in the correlation function $C$,
due to
phase space limitations and charge conservation constraints.
In addition, the choice of unlike-sign pairs as the reference sample
adds further distortions to the correlation function, due to pions from
resonance decays.
To reduce these effects, we introduced the (double) ratio $C'$ of
the correlation functions $C$ in the data and in a sample of 7.2 million
Jetset 7.4~\cite{jetset} multihadronic Monte Carlo (MC) events, without BECs:
\begin{equation}
C'=\frac{C^{\rm DATA}}{C^{\rm MC}}=
\frac{N^{\rm DATA}_{\rm like}/N^{\rm DATA}_{\rm unlike}}
{N^{\rm MC}_{\rm like}/N^{\rm MC}_{\rm unlike}}.
\end{equation}
The Monte Carlo samples are processed through a full simulation
of the OPAL detector \cite{MCsim}. The simulation parameters
of the generator were tuned in \cite{MCtun}.
\par
\begin{figure}[t]
\centerline{\epsfxsize=10cm\epsffile{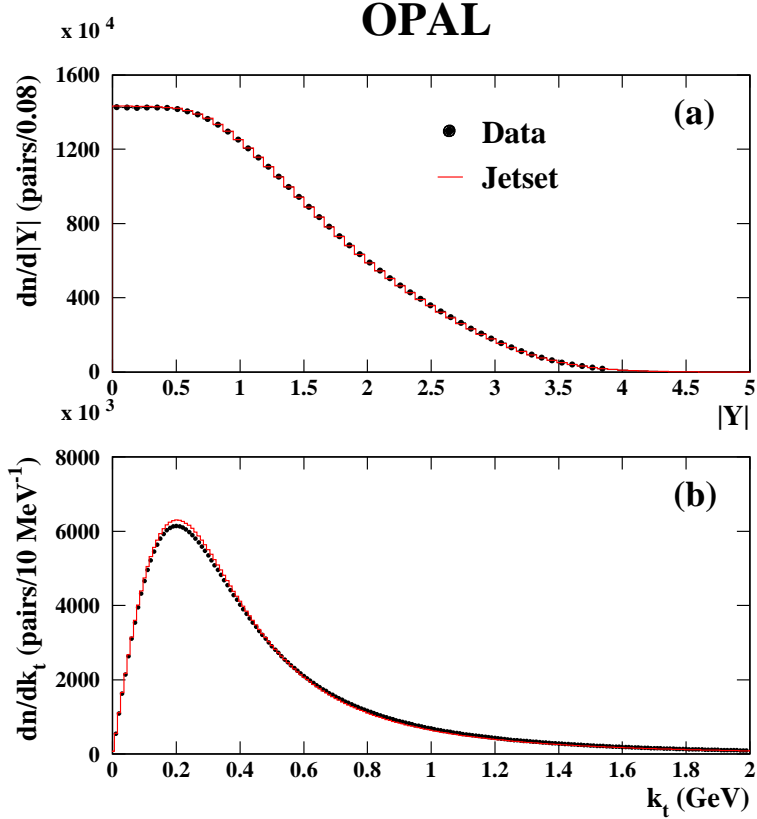}}
\caption{\sl{(a) Histogram of the differential distribution in
             the pair rapidity $|Y|$ and (b) in the pair mean transverse
             momentum $\kt$ of the data (dots) and Jetset events
             (line).
             The number of pairs in the Monte Carlo sample has been normalized
             to the number of pairs in the data sample.}}
\label{ykt}
\end{figure}

The dependence of the correlation functions
$C'(\qt,\ql,\qz)$ and $C'(\Ql,\Qts,\Qto)$
on the pair average 4-momentum $K$ has been analyzed by selecting pions
in different intervals of two components of $K$: the pair rapidity
\begin{equation}
|Y| = \frac{1}{2} \ln \left[
{\frac{(E_1+E_2)+(p_{{\mathrm \ell},1}+p_{{\mathrm \ell},2})}
{(E_1+E_2)-(p_{{\mathrm \ell},1}+p_{{\mathrm \ell},2})}}\right]
\end{equation}
and the pair average transverse momentum with respect to the event thrust
direction
\begin{equation}
\kt = \frac{1}{2} \left|
(\vec{p}_{{\mathrm t},1} + \vec{p}_{{\mathrm t},2}) \right|.
\end{equation}
The differential $|Y|$ and $\kt$ distributions,
$\frac{dn}{d|Y|}$ and $\frac{dn}{dk_{\mathrm t}}$,
of the data are shown in Fig.~\ref{ykt}.
The same distributions for Jetset events are also presented in
Fig.~\ref{ykt}: the comparison shows a good agreement between data
and Monte Carlo events.
\par
The dependence of $C$ and $C'$ on $K$ has been
studied in three bins of $|Y|$
($0.0\leq|Y|<0.8$, $0.8\leq|Y|<1.6$, $1.6\leq|Y|<2.4$)
and five bins of $\kt$ ($0.1\leq\kt<0.2$ GeV,
$0.2\leq\kt<0.3$ GeV, $0.3\leq\kt<0.4$ GeV,
$0.4\leq\kt<0.5$ GeV, $0.5\leq\kt<0.6$ GeV).
In this domain, a total of 47.3 million like-charge and 54.7 million
unlike-charge pairs have been analysed.
%
%
\section{The experimental correlation functions}
\begin{figure}[t]
\centerline{\epsfxsize=12cm\epsffile{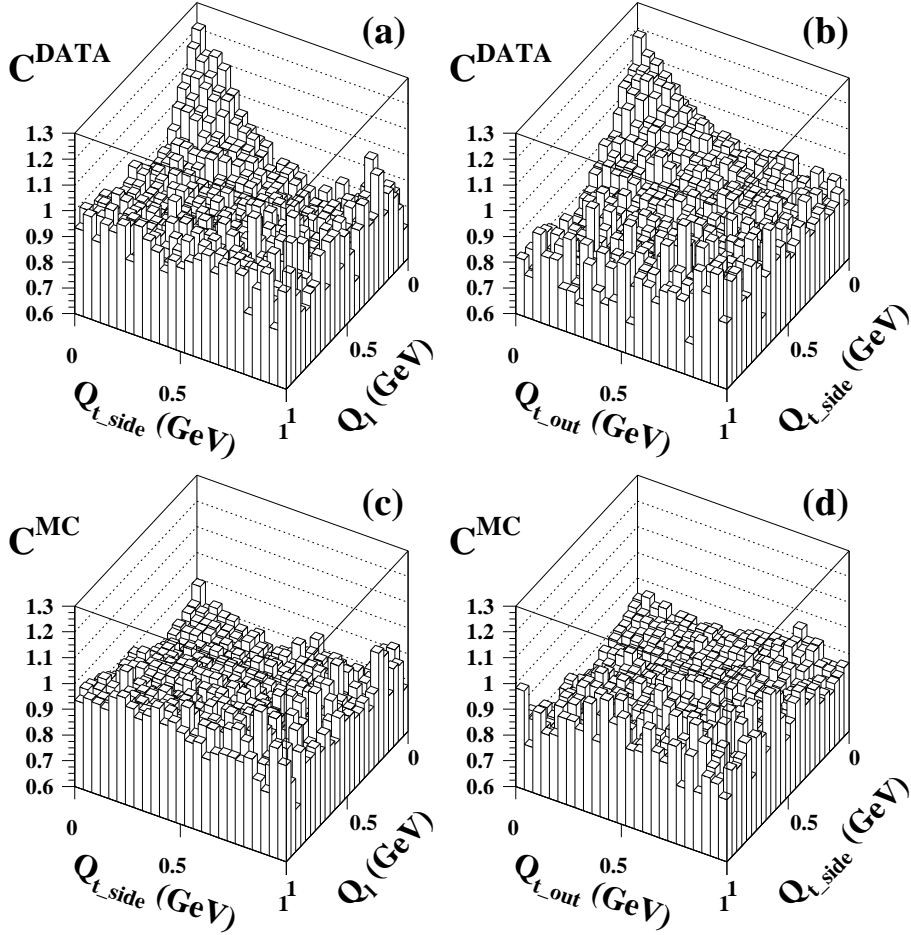}}
\caption{\sl{Two-dimensional projections of the correlation
             function $C(\Ql,\Qts,\Qto)$
             for 0.8 $\leq |Y| <$ 1.6 and
             0.3 GeV $\leq \kt <$ 0.4 GeV for the data
             ((a) and (b)) and for Jetset MC events ((c) and (d)).
             $\Qto <$ 0.2 GeV in (a) and (c);
             $\Ql <$ 0.2 GeV in (b) and (d).
             }}
\label{proBP_1}
\end{figure}

\begin{figure}[t]
\centerline{\epsfxsize=12cm\epsffile{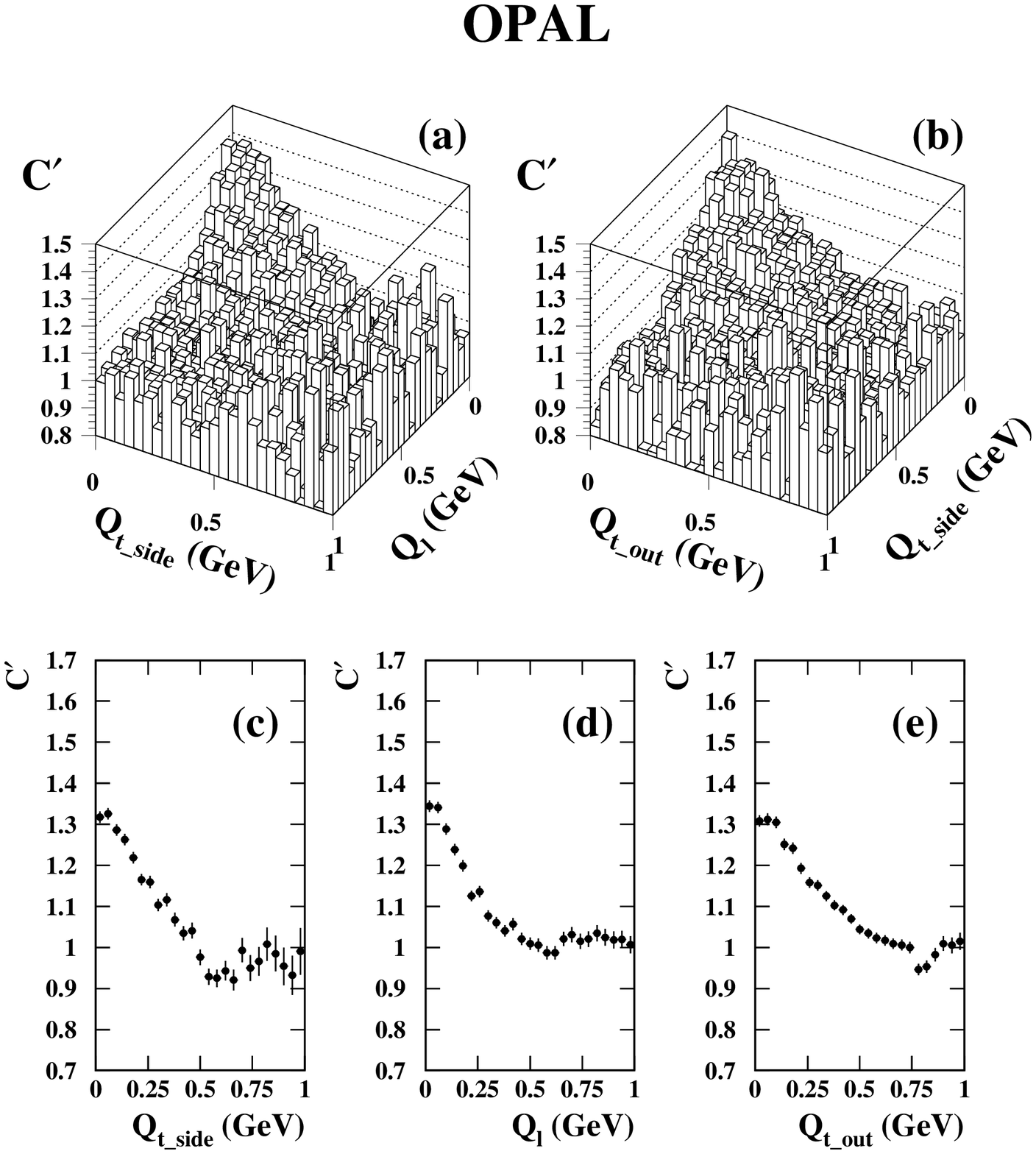}}
\caption{\sl{Two-dimensional ((a) and (b)) and one-dimensional ((c), (d) and
             (e)) projections of the correlation function
             $C'(\Ql,\Qts,\Qto)$
             for 0.8 $\leq |Y| <$ 1.6 and
             0.3 GeV $\leq \kt <$ 0.4 GeV.
             $\Qto <$ 0.2 GeV in (a),
             $\Ql <$ 0.2 GeV in (b).  In (c), (d) and (e)
             the one-dimensional projections are obtained for low values
             ($<$ 0.2 GeV) of the remaining two variables.
             }}
\label{proBP_2}
\end{figure}

%
%
Samples of two-dimensional projections of the correlation function 
$C(\Ql,\Qts,\Qto)$ for a single bin of $|Y|$ and $\kt$ are shown
in Fig.~\ref{proBP_1} for the data and the MC Jetset events.
For the example shown~\footnote{Files of the three-dimensional correlation
functions will be made available in the Durham HEP database.}, the bin
corresponding to pair rapidities and transverse
momenta in the intervals $0.8 \leq |Y| < 1.6$ and
0.3 GeV $\leq \kt< 0.4$ GeV was chosen.
Small ($< 0.2$ GeV) values of $\Qto$ and of
$\Ql$ have been required
in the ($\Ql,\Qts$) and in the
($\Qts,\Qto$) projections, respectively.
Bose-Einstein correlation peaks are visible in the data at low
$\Ql,\Qts,\Qto$
but they are not present in the Monte Carlo samples.
The same two-dimensional projections for the correlation function
$C'(\Ql,\Qts,\Qto)$
are presented in Fig.~\ref{proBP_2}(a) and (b).
Also shown, in Fig.~\ref{proBP_2}(c), (d) and (e), are the one-dimensional
projections for low ($<0.2$ GeV) values of the other two variables.
\par
\begin{figure}[t]
\centerline{\epsfxsize=12cm\epsffile{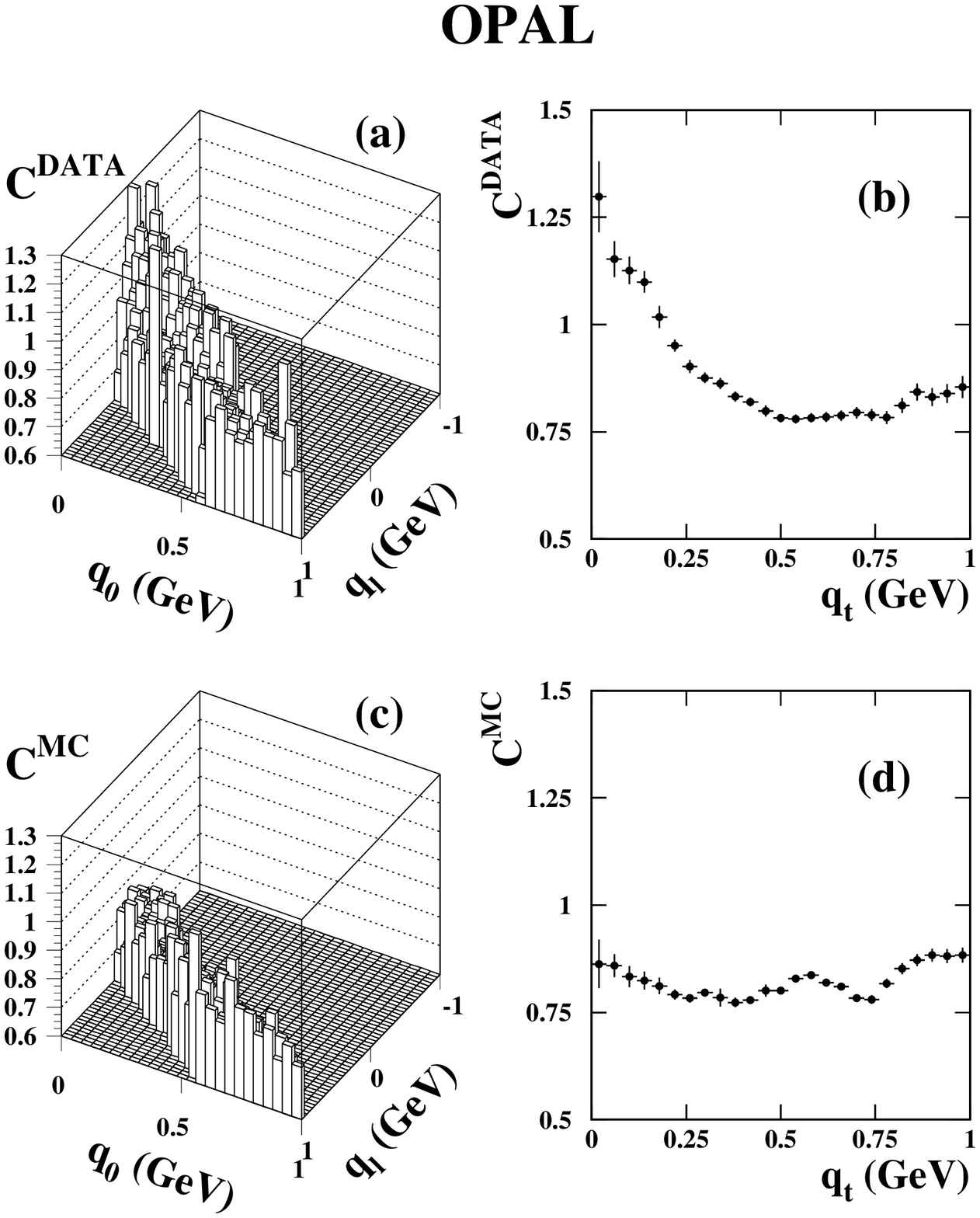}}
\caption{\sl{Two-dimensional ($\ql,\qz$) and one-dimensional $\qt$
             projections of the correlation
             function $C(\qt,\ql,\qz)$
             for data ((a) and (b)) and Jetset events ((c) and (d)).
             The correlation function was measured in the bin
             0.8 $\leq |Y| <$ 1.6 and
             0.3 GeV $\leq \kt <$ 0.4 GeV.
             It was required $\qt <$ 0.2 GeV in (a) and (c).
             In (b) and (d) the one-dimensional projections are obtained
             for low values ($<$ 0.2 GeV) of the remaining two variables.
             }}
\label{proYK}
\end{figure}
\begin{figure}[t]
\centerline{\epsfxsize=12cm\epsffile{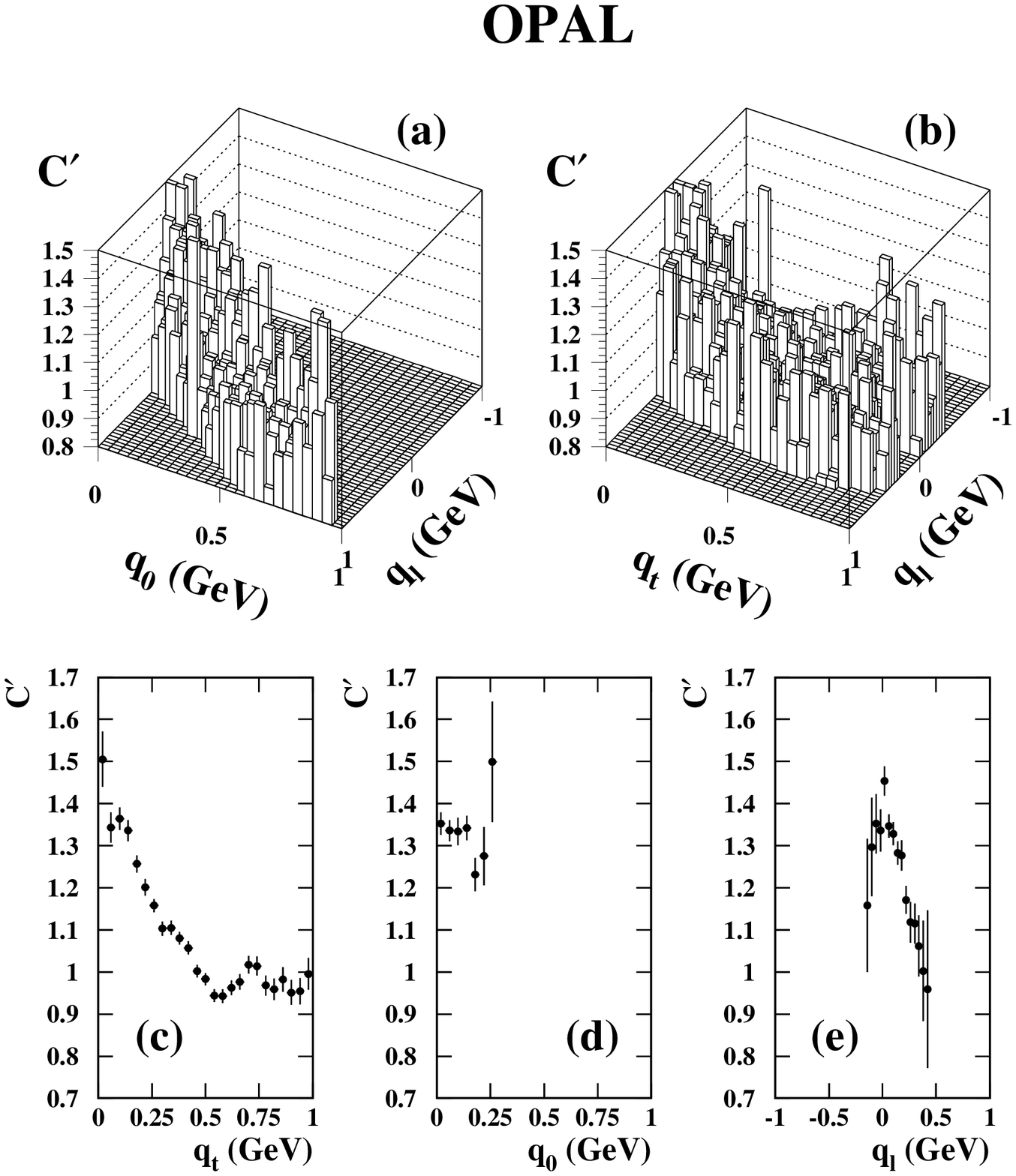}}
\caption{\sl{Two-dimensional projections of the correlation
             function $C'(\qt,\ql,\qz)$: ($\ql,\qz$)
             for $\qt<$ 0.2 GeV in (a) and
             ($\ql,\qt$) for $\qz<$ 0.2 GeV in (b).
             One-dimensional projections ((c), (d) and (e)) of
             $C'(\qt,\ql,\qz)$, obtained
             for low values ($<$ 0.2 GeV) of the remaining two variables.
             The correlation function has been measured in the bin
             0.8 $\leq|Y|<$ 1.6 and
             0.3 GeV $\leq \kt <$ 0.4 GeV.
             }}
\label{proYK1}
\end{figure}

The two-dimensional ($\ql,\qz$) and the one-dimensional $\qt$
projections of the correlation function
$C(\qt,\ql,\qz)$
in the bin 0.8 $\leq|Y|<$ 1.6 and 0.3 GeV $\leq\kt<$ 0.4 GeV
are shown in Fig.~\ref{proYK}, for data and Jetset events.
Narrow cuts ($<0.2$ GeV) on the other variables have been applied to make
the projections.
The combination [($\qt^2+\ql^2)-\qz^2$] of the three variables is an
invariant greater than zero.
This condition and the bound on the pair rapidity constrain the
correlation function to be different from zero only in a limited region of
the ($\ql,\qz$) plane, as can be seen in Fig.~\ref{proYK}~(a) and (c).
The ($\ql,\qz$) and ($\ql,\qt$) projections of the correlation
function $C'(\qt,\ql,\qz)$ are
shown in Fig.~\ref{proYK1} together with the one-dimensional projections,
for small ($<0.2$ GeV) values of the other variables.
BEC enhancements are clearly seen in both the $\qt$ and $\ql$
projections, Fig.~\ref{proYK1}~(c) and (e).
Fig.~\ref{proYK1}~(d), on the other hand, shows that the range available
to the variable $\qz$ is quite restricted, and that no Bose-Einstein peak
can be observed.
%
%
\section{Parameterizations of the correlation functions}
%
%
To extract the spatial and temporal extensions of the pion source from
the experimental correlation functions,
the Bertsch-Pratt (BP)~\cite{bertsch_pratt}
\begin{flushleft}
$C'(\Ql,\Qts,\Qto)$
\end{flushleft}
\begin{equation}
= N (1 + \lambda {\rm e}^{-(\Ql^2 \Rlong^2 +
\Qts^2 \Rts^2 +
\Qto^2 \Rto^2 +
2 \Ql \Qto \Rlongto^2
)})
F(\Ql,\Qts,\Qto)
\end{equation}
and the Yano-Koonin (YK)~\cite{yano_koonin}
\begin{equation}
C'(\qt,\ql,\qz) =
N (1 + \lambda {\rm e}^{-(\qt^2 \Rt^2 +
\gamma^2(\ql - v \qz)^2 \Rl^2 +
\gamma^2(\qz - v \ql)^2 \Rz^2
)})
F(\qt,\ql,\qz)
\end{equation}
parameterizations were fitted to the measured correlation
functions in all intervals of $\kt$ and $|Y|$.
\par
In both parameterizations, $N$ is a normalization factor while
$\lambda$ measures the degree of incoherence of the pion sources, and is
related to the fraction of pairs that actually interfere.
The two parameters $N$ and $\lambda$, whose product determines the size of the
BEC peak, are however significantly (anti)correlated:
this limits the interpretation of $\lambda$ and the comparison of its values
between the two parameterizations.
\par
The two functions $F(\Ql,\Qts,\Qto) =
(1 + \elong\Ql + \ets\Qts + \eto\Qto)$
and $F(\qt,\ql,\qz) = (1 + \dt\qt + \dl\ql + \dz\qz)$,
where $\epsilon_{\mathrm i}$ and $\delta_{\mathrm i}$ are free
parameters, were introduced in Eq.~(7) and (8) to take into
account residual long-range two-particle correlations, due to energy and
charge conservation.
\par
The interpretation of the other free parameters in Eq~(7), is the
following:
\begin{itemize}
\item $\Rts$ and $\Rlong$ are the transverse and longitudinal
source radii in the LCMS, i.e. the longitudinal rest frame of
the pair;
\item $\Rto$ and the cross-term $\Rlongto$ are a
combination of both the spatial and temporal extentions of the source.
Under certain assumptions~\cite{time_rel}, the difference
($\Rto^2 - \Rts^2$) is proportional to the duration of
the particle emission process, and $\Rlongto$ to the source
velocity with respect to the pair rest frame~\cite{cross_term}.
\end{itemize}
In the YK function Eq.~(8), where $\gamma=1/\sqrt{1-v^2}$, the free
parameters are interpreted as follows:
\begin{itemize}
\item $v$ is the longitudinal velocity, in units of $c$,
of the source element in the CMS frame;
\item $\Rz$ measures the time interval, times $c$, during which
particles are emitted, in the rest frame of the emitter (source element).
Difficulties in achieving reliable results for the time parameter
$\Rz^2$ in YK fits have been reported in the literature~\cite{timepar},
due to the limited phase-space available in
$\gamma^2(\qz-v\ql)^2$;
\item $\Rt$ and $\Rl$ are the transverse and longitudinal radii,
i.e. the regions of homogeneity of the source, in the rest frame of the
emitter.
\end{itemize}
\par
The parameters $\Rz$, $\Rt$ and $\Rl$ do not depend on the
frame in which the correlation function has been measured, since they are
evaluated in the rest frame of the source element.
\par
The two parameterizations are not independent~\cite{time_rel}, so that a
comparison between the BP and the YK fits
represents an important test.
%
%
\section{Results}
%
%
Minimum $\chi^2$ fits of the Bertsch-Pratt and the Yano-Koonin
parameterizations to the experimental correlation functions were performed
using the MINUIT~\cite{minuit} program.
The error associated to each entry of the three-dimensional matrices
$C$ and $C'$ was computed attributing a Poissonian uncertainty to
the number of like and unlike charge pairs in the corresponding bin.
The fit range allowed to each variable was set between 40 MeV and 1 GeV.
The region below 40 MeV was excluded to avoid problems of detector
resolution and poorly reconstructed or split tracks which mimic two like
charged particle tracks with very low $q$.
In Sections 5.1 and 5.2 the results of the fits are presented.
Sources of systematic uncertainties on the fit parameters are discussed
in Section 5.3.
Section 5.4 is devoted to a comparison between the BP and the YK
parameterizations.
%
%
\subsection{Bertsch-Pratt fits}
\begin{figure}[t]
\centerline{\epsfxsize=14cm\epsffile{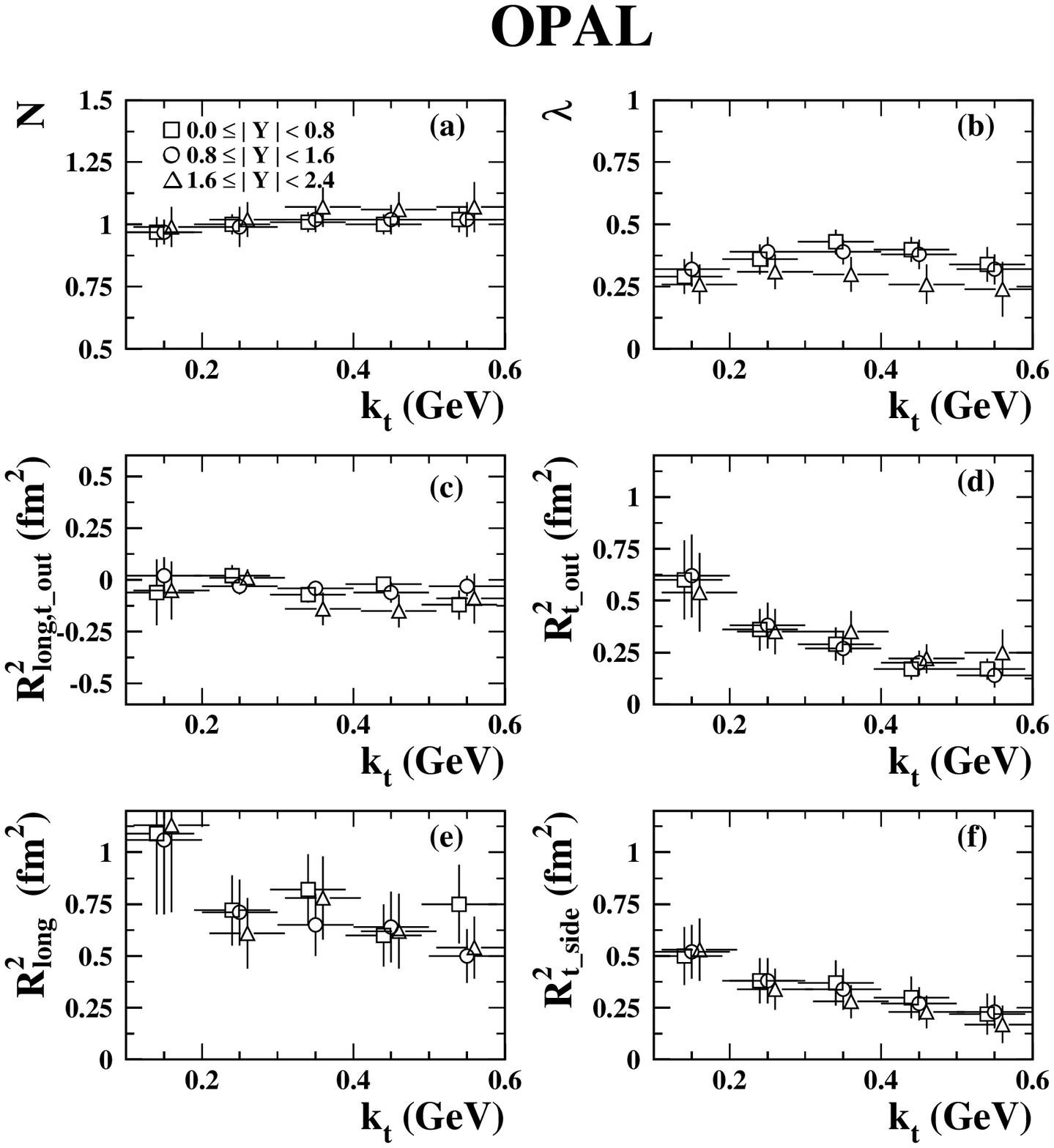}}
\caption{\sl{Best-fit parameters of the Bertsch-Pratt parameterization,
             Eq.~(7), to the correlation function
             $C'(\Ql,\Qts,\Qto)$,
             as a function of $\kt$, for different
             intervals of rapidity $|Y|$.
             The correlation functions were measured in the LCMS frame.
             Horizontal bars represent bin widths and vertical bars
             include both statistical and systematic errors.
             (a) the normalization factor N;
             (b) the incoherence parameter $\lambda$;
             (c) the cross term $\Rlongto^{2}$;
             (d) the parameter $\Rto^2$;
             (e) the squared longitudinal correlation length
             $\Rlong^2$ and
             (f) the squared transverse correlation length
             $\Rts^2$.}}
\label{all9parbp}
\end{figure}

%
%
The best-fit parameters of the BP function, Eq.~(7), are listed in
Table~\ref{bp}, and their dependence on $|Y|$ and $\kt$ is shown
in Fig.~\ref{all9parbp}.
Errors in Fig.~\ref{all9parbp} include both
statistical standard deviations as given by the fit program~\footnote{The
HESSE algorithm in MINUIT calculates the error matrix inverting the matrix
of the second derivatives of the fit function with respect to the fit
parameters.} and systematic uncertainties (discussed in Section 5.3), added
in quadrature.
One notes that there is only a minor dependence on the rapidity, but some
parameters depend on $\kt$.
In more detail:
\begin{itemize}
\item $\lambda$ varies between 0.25 and 0.4.
The coefficient of correlation between the parameters $\lambda$ and $N$ is
about $-0.35$, almost independent of $\kt$;
\item $\Rts^2$, $\Rto^2$ and, less markedly,
$\Rlong^2$ decrease with increasing $\kt$.
The presence of correlations between the particle production points
and their momenta is an indication that the pion source is not static, but
rather expands during the particle emission process.
$\Rlong^2$ is larger than the corresponding transverse parameter
$\Rts^2$, in agreement with a pion source which is elongated in
the direction of the event thrust axis~\cite{tr_long};
\item the cross-term parameter $\Rlongto^2$ is
compatible with zero, apart from a few bins at the highest rapidity interval.
This result may be explained~\cite{time_rel} assuming that the source
velocity, measured with respect to the rest frame of the pion pair, is close
to zero;
\item the difference between the ``out" and ``side" transverse parameters,
$(\Rto^2-\Rts^2)$
for $|Y|<$ 1.6 is positive at low $\kt$, then
it decreases and becomes negative for $\kt \geq$ 0.3 GeV.
In the highest rapidity interval, 1.6 $\leq|Y|<$ 2.4,
$(\Rto^2-\Rts^2)$ is compatible
with zero, for all $\kt$.
As a consequence, it is not possible to estimate the particle emission time
from $(\Rto^{2}-\Rts^2)$;
\item the parameters $\epsilon_{\mathrm i}$ are not negligible: the
function $F(\Ql,\Qts,\Qto)$
typically differs from unity for at most 15\% to 20\%
at $Q_{\mathrm i} \approx 1$ GeV.
\end{itemize}
%
%
\subsection{Yano-Koonin fits}
\begin{figure}[t]
\centerline{\epsfxsize=14cm\epsffile{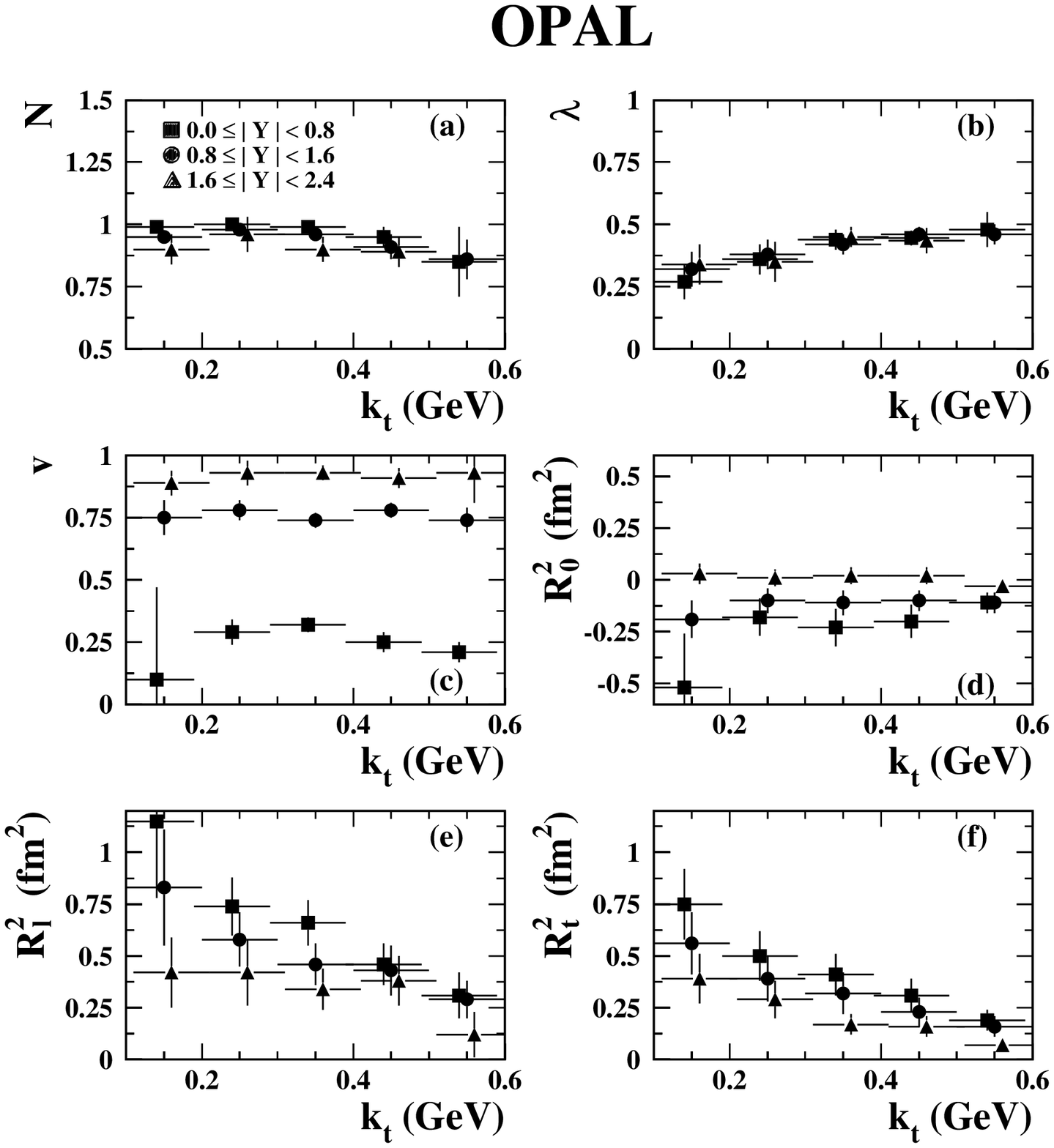}}
\caption{\sl{Best-fit parameters of the Yano-Koonin parameterization,
             Eq.~(8), to the correlation function $C'(\qt,\ql,\qz)$,
             as a function of $\kt$, for different intervals of
             rapidity $|Y|$.
             The correlation functions were measured in the event
             centre-of-mass frame.
             Horizontal bars represent bin widths and vertical bars
             include both statistical and systematic errors.
             (a) the normalization factor N;
             (b) the parameter $\lambda$;
             (c) the source velocity $v$;
             (d) the time parameter $\Rz^2$;
             (e) the squared longitudinal correlation length
             $\Rl^2$ and
             (f) the squared transverse correlation length
             $\Rt^2$.}}
\label{all9par}
\end{figure}

\begin{figure}[t]
\centerline{\epsfxsize=10cm\epsffile{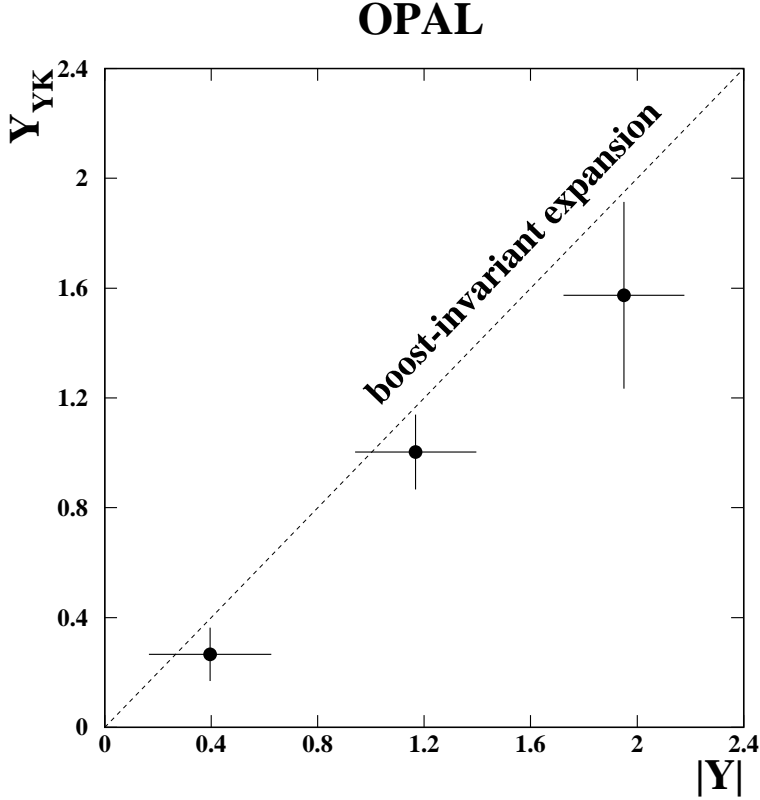}}
\caption{\sl{The Yano-Koonin rapidity $\YYK$ plotted versus
             the pion pair rapidity $|Y|$.
             Each $|Y|$ was computed as the weighted
             average of the corresponding bin.
             $\YYK$ values were computed by means of Eq.~(9), using
             the average value of $v$ over all $\kt$ in that
             $|Y|$ bin.
             Horizontal bars are r.m.s. deviations
             from the average.
             Vertical bars include both statistical and systematic
             errors.
             Also shown is the line $\YYK$=$|Y|$, corresponding to
             a source which expands boost-invariantly.}}
\label{yy}
\end{figure}

%
%
Table~\ref{yk} and Fig.~\ref{all9par} show the parameters of the YK fits,
Eq.~(8), in different $|Y|$ and $\kt$ intervals.
Error bars in Fig.~\ref{all9par} include both statistical and systematic
uncertainties, added in quadrature.
It can be seen that:
\begin{itemize}
\item the parameter $\lambda$ is almost independent of rapidity and increases
with $\kt$, reaching values of about 0.5 for the largest $\kt$
values. It is however significantly anticorrelated with the parameter $N$,
the correlation coefficient increasing in absolute value from about
$-0.50$ at low $\kt$ up to $-0.80$ for $\kt>0.4$ GeV;
\item both $\Rt^2$ and $\Rl^2$ decrease with increasing $\kt$ and $|Y|$.
The longitudinal radii are larger than the transverse radii.
This agrees with an expanding, longitudinally elongated source;
\item $\Rz^2$ is compatible with zero at high rapidities, and assumes
negative values for $|Y|<1.6$.
This excludes an interpretation of $\Rz/c$ in terms of
the time duration of the particle emission process;
\item those of the parameters $\delta_{\mathrm i}$ which are not negligible,
contribute typically 10\% to 15\% to the function
$F(\qt,\ql,\qz)$ at large $q_{\mathrm i}$;
\item the source velocity $v$ does not depend on $\kt$, but it
is strongly correlated with the pair rapidity.
\end{itemize}
The dependence of $v$ on $|Y|$ can also be presented~\cite{gibs,h_ions}
in terms of a plot \`{a} la GIBS, i.e. the Yano-Koonin rapidity
\begin{equation}
\YYK=\frac{1}{2} \ln \left(\frac{1+v}{1-v}\right)
\end{equation}
as a function of the pair rapidity $|Y|$.
$\YYK$ measures the rapidity of the source element with respect to the
centre-of-mass frame:
a non-expanding source would therefore correspond to $\YYK\approx0$
for any $|Y|$.
On the other hand, for a boost-invariant source~\footnote{
A source expands boost-invariantly in the longitudinal direction if
the velocity of each element is given by $v=z/t$,
where $t$ and $z$ are, respectively, the time elapsed since the collision
and the longitudinal coordinate of the element, in the centre-of-mass frame.
In that case, particle emission happens at constant proper times
$\sqrt{t^2-z^2}$.}, the strict
correlation $\YYK=|Y|$ is expected~\cite{time_rel,yano_koonin}, since only
the
source elements which move with velocities close to the velocity of the
observed particle pair contribute to the correlation function.
In Fig.~\ref{yy} the Yano-Koonin rapidity $\YYK$ is shown as a function
of the pair rapidity.
Since in a given $|Y|$ interval the parameter $v$ is almost independent of
$\kt$, see Fig.~\ref{all9par}(c), each $\YYK$ is computed, according to
Eq.~(9), using the average value of $v$ over all $\kt$ in that $|Y|$ bin.
Each $|Y|$ has been computed as the weighted average of the
corresponding bin, rather than the centre of the bin.
A clear positive correlation between $\YYK$ and $|Y|$ is observed, even if
$\YYK<|Y|$ at the largest pair rapidities.
This is in agreement with a pion source which is emitting particles in a
nearly boost-invariant way.
\par
\begin{figure}[t]
\centerline{\epsfxsize=12cm\epsffile{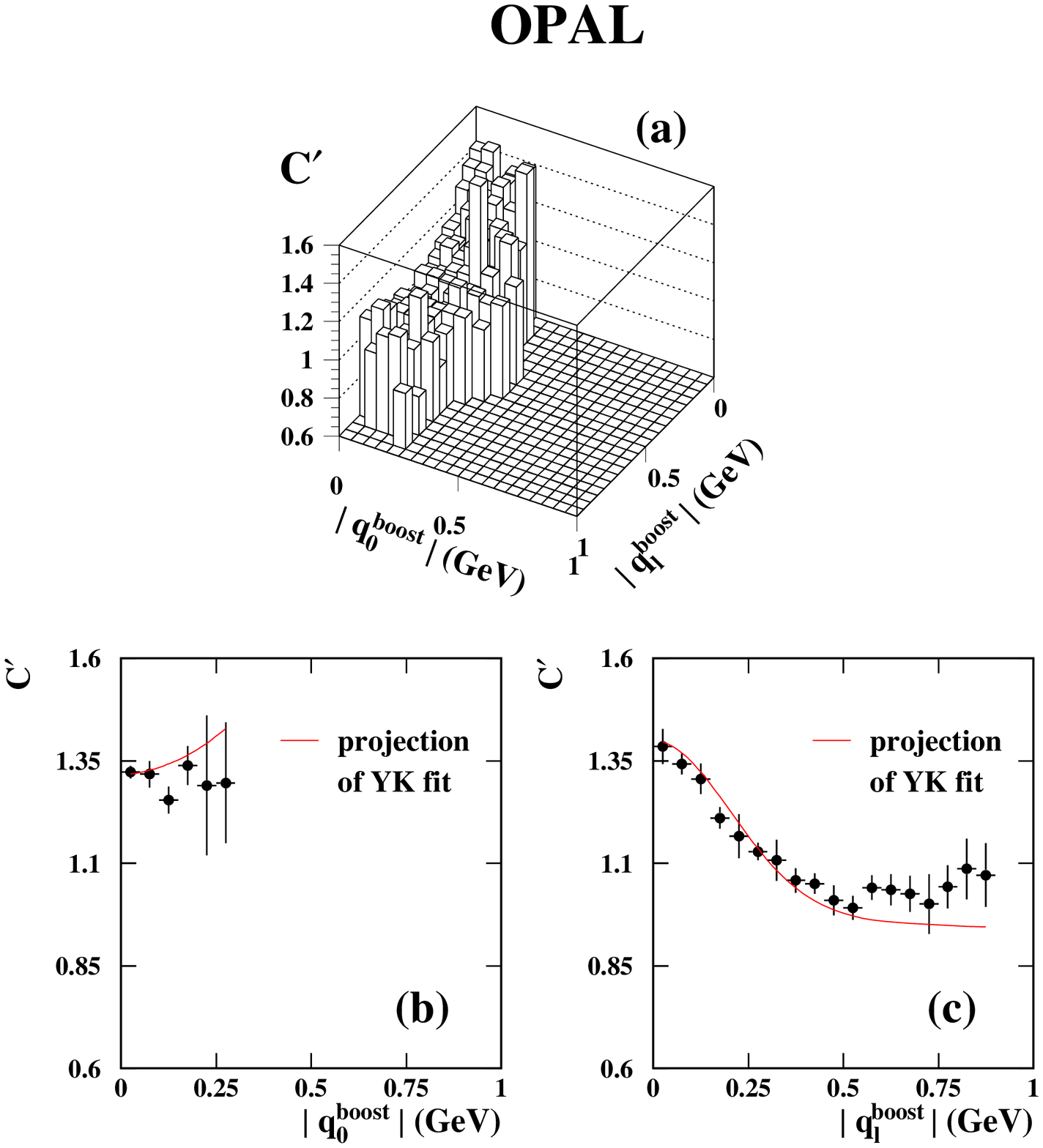}}
\caption{\sl{(a) The two-dimensional projection
             ($|\ql^{\mathrm {boost}}|,|\qz^{\mathrm {boost}}|$),
             after the longitudinal boost to the
             source element rest frame,
             measured for pion pairs
             in the rapidity interval 0.8 $\leq |Y| <$ 1.6 and with mean
             transverse momenta in the range
             0.3 GeV $\leq \kt <$ 0.4 GeV.
             The projection was made requiring
             $\qt <$ 0.2 GeV.\newline
             (b) The one-dimensional projection in $|\qz^{\mathrm {boost}}|$
             ($|\ql^{\mathrm {boost}}|<$ 0.2 GeV). The curve
             is the one-dimensional projection of the
             Yano-Koonin three-dimensional best-fit function.\newline
             (c) The one-dimensional projection in $|\ql^{\mathrm {boost}}|$
             ($|\qz^{\mathrm {boost}}|<$ 0.2 GeV). The curve
             is the one-dimensional projection of the
             Yano-Koonin three-dimensional best-fit function.}}
\label{pr0l}
\end{figure}

To try to understand the YK fit results of the parameter $\Rz^2$,
it is useful to analyse the two-dimensional projection ($\ql,\qz$)
of the correlation function
$C'(\qt,\ql,\qz)$
after the longitudinal boost to the rest frame of the source element.
We then introduce
$\ql^{\mathrm {boost}}=\gamma(\ql-v\qz)$ and
$\qz^{\mathrm {boost}}=\gamma(\qz-v\ql)$, where
the best-fit parameter $v$ is used to boost the variables.
In Fig.~\ref{pr0l}(a) the two-dimensional
($|\ql^{\mathrm {boost}}|$,$|\qz^{\mathrm {boost}}|$)
projection of $C'$ is presented.
The phase space available to $|\qz^{\mathrm {boost}}|$ is limited, when $\qt$
approaches 0, and the one-dimensional $|\qz^{\mathrm {boost}}|$ projection
(Fig.~\ref{pr0l}(b)) is approximately flat: it is not possible to
distinguish any peak due to Bose-Einstein correlations and, for most rapidity
and $\kt$ intervals, the fitted $\Rz^2$ have negative values. 
In Fig.~\ref{pr0l}(b), the solid line shows the one-dimensional
$|\qz^{\mathrm {boost}}|$ projection ($|\ql^{\mathrm {boost}}|<0.2$ GeV,
$\qt<0.2$ GeV) of the YK fit, Eq.~(8);
the line is an increasing function of $\qz^{\mathrm {boost}}$, because of the
negative value of $\Rz^2$.
Similar limitations in the temporal acceptance
have been reported in the literature~\cite{timepar}.
On the other hand, the $|\ql^{\mathrm {boost}}|$ projection for
$|\qz^{\mathrm {boost}}|<0.2$ GeV and $\qt<0.2$ GeV,
Fig.~\ref{pr0l}(c), shows a clear BEC peak at
small $|\ql^{\mathrm {boost}}|$, reproduced by the one dimensional
$|\ql^{\mathrm {boost}}|$
projection of the best-fit YK function (solid line).
%
%
\subsection{Systematic effects}
%
%
The systematic uncertainties of the fit parameters and
the stability of the results concerning the dependence
of the transverse and longitudinal radii on $\kt$
was studied by considering a number of changes with respect
to the reference analysis.
The following changes were taken into account:
\begin{itemize}
\item A correction was applied to the correlation functions, based
on the Gamow factors~\cite{gamow}, in order to take into account
final-state Coulomb interactions between charged pions.
\item The analysis was repeated with more stringent cuts in the
selection: a maximum momentum of 30 GeV instead
of 40 GeV and a charge unbalance smaller than 0.25 per event instead of 0.4.
\item The fits were repeated changing the upper bound of the fit range
from 1 GeV to 0.8 GeV.
\end{itemize}
In the cases listed above, we found negligible differences in the parameters
with respect to the reference analysis.
The systematic effect on the correlation function $C'$, due to
the Monte Carlo modelling, was assumed negligible.
\begin{itemize}
\item The correlation functions were measured in bins of 60 MeV,
instead of 40 MeV, to test the stability of the fits.
Bin widths larger than 60 MeV would prevent a correct reconstruction of the
BEC peak, which is about 300$\div$400 MeV wide.
\item Possible non-Gaussian shapes of the correlation functions at
low $q$ were tested replacing the Gaussian functions in the BP and YK
parameterizations with first order Edgeworth expansions~\cite{edgeworth}
of the Gaussian.
The $\chi^2$/DoF of the two fits were found to be comparable.
\end{itemize}
Systematic errors on the fit parameters have been computed adding in
quadrature the deviations from the standard fit; they are reported in
Tables~\ref{bp} and~\ref{yk}.
\par
Assuming simple linear dependences of the squared BP and YK longitudinal and
transverse radii on $\kt$, we measured the slopes, $dR_{\mathrm i}^2/d\kt$, by
minimum $\chi^2$ fits.
Fits were performed on the radii of the reference analysis, with statistical
errors only.
The systematic errors on the slopes were then estimated comparing the
slopes from the reference analysis with the slopes from the systematic
checks listed above.
Table~\ref{slope} shows the best-fit slopes with errors.
In all cases a decrease of the
radii with increasing $\kt$ is favoured even if, in one rapidity interval,
the longitudinal BP radius is compatible with independence on $\kt$.
\par
To investigate further the decrease of the radii on $\kt$, the YK and BP
functions were fitted to the correlation function $C$, Eq.~(1).
Larger (about 30\%) squared transverse and longitudinal radii with respect
to the correlation function $C'$ are obtained in this case.
However, the slopes of the linear dependences of the squared radii on
$\kt$ are the same, within uncertainties, for $C$ and $C'$.
A comparison of the YK best-fit parameters from minimizing 
$\chi^2$ values and from maximizing a likelihood function~\cite{e802} has
been done for the correlation function $C$.
The differences between the parameters fitted with the two techniques were
negligible.
\par
One more check was done on the YK transverse radius $\Rt$:
we computed the one-dimensional projection
$C'(\qt,0,0)$ of the three-dimensional correlation function
$C'(\qt,\ql,\qz)$, by requiring $\ql$ and
$\qz\leq0.08$ GeV, and we fitted the function
\begin{equation}
C'(\qt) = N (1+\lambda {\rm e}^{-\qt^2 \Rt^2})
\end{equation}
to the projection.
We first checked that the best-fit $\Rt^2$ is compatible, within
errors, to the one we obtain if the right-hand side of Eq.~(10) is multiplied
by a ``long-range" factor ($1+\dt\qt$).
Based on the same one-dimensional projection $C'(\qt,0,0)$,
we also measured the transverse correlation length in a fit-independent
way~\cite{heinz}, introducing the parameter $\Rt$
\begin{equation}
\tilde{R}_{\mathrm t}=\frac{1}{\sqrt{2\langle\qt^2\rangle}}~~~~
{\mathrm {where}}~~~~\langle\qt^2\rangle=\frac{\int \qt^2
[C'(\qt,0,0)-1]d\qt}
{\int [C'(\qt,0,0)-1]d\qt}
\end{equation}
i.e. the inverse variance of the correlation function for small $\qt$
values~\footnote{In the actual estimate of $\langle \qt^2 \rangle$ we
have computed
$\frac{\sum \qt^2[C'(\qt,0,0)-N]}
{\sum [C'(\qt,0,0)-N]}$,
where $N$ is the normalization parameter of the fit Eq.~(10) and each $\qt$
has been taken as the central value of the corresponding 40 MeV bin.}.
We found that $\Rt$, computed using Eq.~(11), agrees with the
best-fit $\Rt$ from Eq.~(10); the slope of the linear decrease
is about 20\% smaller than the one measured with three-dimensional YK fits,
Eq.~(8).
\par
The standard analysis was also repeated for a subsample of events classified
as two-jets by the Durham jet-finding algorithm~\cite{durham}.
The subsample was defined by setting the resolution parameter at
$y_{\mathrm {cut}}=0.04$.
The dependences of the best-fit parameters on $|Y|$ and $\kt$ are similar
to those found for the inclusive sample of events.
In particular, the longitudinal and the transverse radii decrease with
increasing $\kt$. However, the radii measured
in the case of two-jet events are smaller, by about 10\%, than in the
inclusive sample~\cite{fierro}.
An increase of the ``jettyness" of the two-jet subsample, obtained using
a smaller $y_{\mathrm {cut}}$ ($y_{\mathrm {cut}}=0.02$) in the jet-finding
algorithm, does
not change significantly the behaviour of the parameters.
%
%
\subsection{Comparison between BP and YK fits}
%
%
The following relations should hold between the correlation lengths of the
BP and YK functions measured in the LCMS and CMS frames,
respectively~\cite{time_rel}:
\begin{eqnarray}
\Rts^2 & = & \Rt^2 \\
\Rlong^2 & = &
\gamma_{\mathrm {LCMS}}^2(\Rl^2+\beta_{\mathrm {LCMS}}^2 \Rz^2) \\
(\Rto^2-\Rts^2) & = & \beta_{\mathrm t}^2
\gamma_{\mathrm {LCMS}}^2(\Rz^2+\beta_{\mathrm {LCMS}}^2\Rl^2).
\end{eqnarray}
In Eq.~(13) and (14) $\beta_{\mathrm {LCMS}}$ is the velocity of the
source element
measured in the LCMS, i.e. with respect to the pair longitudinal rest frame;
$\gamma_{\mathrm {LCMS}}=1/\sqrt{1-\beta_{\mathrm {LCMS}}^2}$.
In Eq.~(14) $\beta_{\mathrm t}^2=\left<\frac{2 \kt}{E_1+
E_2}\right>^{2}$, where the brackets stand
for the average over all pion pairs in the given $|Y|$ and $\kt$ range.
For a boost-invariant source, $\beta_{\mathrm {LCMS}}=0$ and Eqs.~(13)
and (14) reduce to:
\begin{eqnarray}
\Rlong^2 & = & \Rl^2 \\
(\Rto^2-\Rts^2) & = & \beta_{\mathrm t}^2 \Rz^2.
\end{eqnarray}
\par
\begin{figure}[t]
\centerline{\epsfxsize=16cm\epsffile{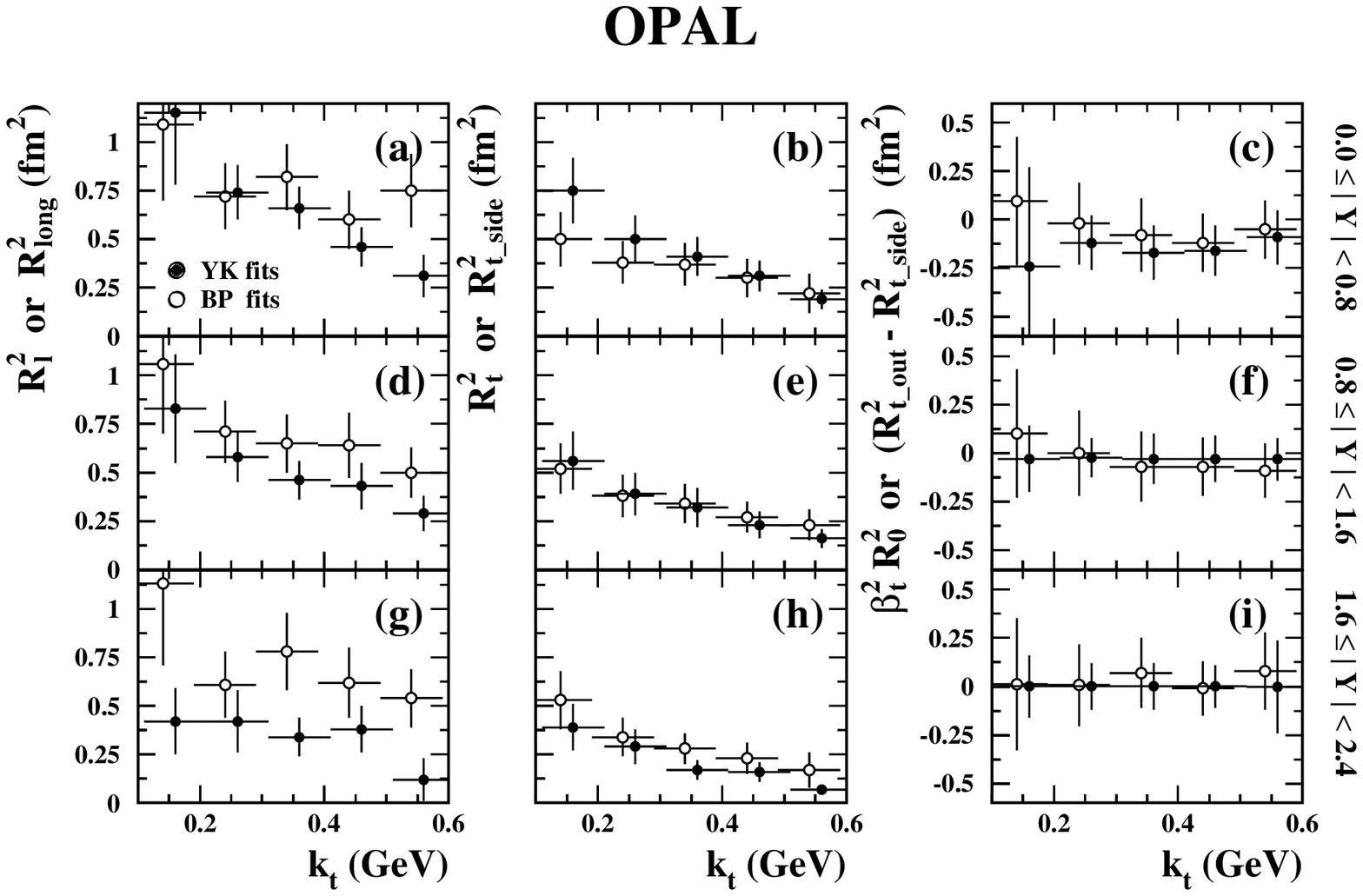}}
\caption{\sl{(a)~(d)~(g) The best-fit longitudinal radius
             $\Rlong^2$ of the Bertsch-Pratt parameterization
             (open dots) compared with the Yano-Koonin
             longitudinal radius $\Rl^2$ (full dots).
             (b)~(e)~(h) The BP transverse correlation length
             $\Rts^2$ (open dots) compared with the YK
             transverse correlation length $\Rt^2$ (full dots).
             (c)~(f)~(i) The difference of the BP transverse radii
             $(\Rto^2-\Rts^2)$
             (open dots) compared with the YK time parameter $\Rz^2$
             times $\beta_{\mathrm t}^2$ (full dots).
             Errors on the parameters include both statistical and
             systematic uncertainties, added in quadrature.}}
\label{all9par_bp}
\end{figure}

In Fig.~\ref{all9par_bp} the best-fit BP parameters
$\Rlong^2$, $\Rts^2$ and $(\Rto^2-\Rts^2)$
are compared with the YK parameters $\Rl^2$, $\Rt^2$ and
$\beta_{\mathrm t}^2$R$_0^2$.
\par
The longitudinal parameter $\Rlong^2$ is systematically larger than
$\Rl^2$ in all the rapidity intervals analyzed
(Fig.~\ref{all9par_bp}(a),~(d)~and~(g)).
According to Eq.~(13), $\Rlong^2 > \Rl^2$
corresponds to $\beta_{\mathrm {LCMS}}$ greater than zero, in agreement with
a pion source whose expansion is not exactly boost-invariant.
\par
The equality of the transverse parameters $\Rts^2$ and $\Rt^2$,
Eq.~(12), is confirmed within errors, with possible
deviations at low $\kt$ (Fig.~\ref{all9par_bp}(b),~(e)~and~(h)).
\par
The negative values of $\Rz^2$ and
$(\Rto^2-\Rts^2)$ appearing in the
two first rapidity intervals (Fig.~\ref{all9par_bp}(c),~(f)~and~(i))
prevent an interpretation in terms of the time duration of the
particle emission process.
Negative values of $\Rz^2$ have been suggested~\cite{opacity} as possible
indicators for opacity of the source, i.e. surface dominated emission.
A dependence of
$(\Rto^2-\Rts^2)$ on $\kt$
similar to the one shown in Fig.~\ref{all9par_bp}(c) and (f) has been
reported in heavy-ion collision experiments~\cite{rhic_puzzle}.
%
%
\section{Conclusions}
%
%
An analysis of Bose-Einstein correlations in \ee annihilation events
at the Z$^0$ peak performed in bins of the average 4-momentum of the pair,
$K$, has been presented for the first time.
Based on this, dynamic features of the pion emitting source were
investigated.
Previous BEC analyses, not differential in $K$, were not sensitive to
these features.
\par
Using the Yano-Koonin and the Bertsch-Pratt formalisms,
the correlation functions were studied in intervals of two components of
$K$: the pion pair rapidity $|Y|$ and the mean transverse momentum $\kt$.
We found that the transverse and longitudinal radii of the pion sources
decrease for increasing
$\kt$, indicating the presence of correlations between the particle
production points and their momenta.
The Yano-Koonin rapidity scales approximately with the pair rapidity,
in agreement with a nearly boost-invariant expansion of the source of pions.
Limitations in the available phase space did not allow measurement of the
duration of the particle emission process.
\par
Similar results have been observed in more complex systems, such as the
pion sources created in pp and heavy-ion collisions, which are now
complemented with such measurements in the simpler hadronic system formed
in \ee annihilations.
\bigskip\bigskip\bigskip\newline
{\Large Acknowledgements}
\bigskip\bigskip\newline
We particularly wish to thank the SL Division for the efficient operation
of the LEP accelerator at all energies
and for their close cooperation with
our experimental group. In addition to the support staff at our own
institutions we are pleased to acknowledge the\\
Department of Energy, USA,\\
National Science Foundation, USA,\\
Particle Physics and Astronomy Research Council, UK,\\
Natural Sciences and Engineering Research Council, Canada,\\
Israel Science Foundation, administered by the Israel
Academy of Science and Humanities,\\
Benoziyo Center for High Energy Physics,\\
Japanese Ministry of Education, Culture, Sports, Science and
Technology (MEXT) and a grant under the MEXT International
Science Research Program,\\
Japanese Society for the Promotion of Science (JSPS),\\
German Israeli Bi-national Science Foundation (GIF),\\
Bundesministerium f\"ur Bildung und Forschung, Germany,\\
National Research Council of Canada,\\
Hungarian Foundation for Scientific Research, OTKA T-038240, 
and T-042864,\\
The NWO/NATO Fund for Scientific Research, the Netherlands.\\

\newpage

\begin{sidewaystable}
\centering
\footnotesize
\begin{tabular}{|l||c|c|c|c|c|}
\hline
{\bf 0.0 $\leq |Y| <$ 0.8}      &
{\bf 0.1 $\leq \kt <$ 0.2 GeV}  &
{\bf 0.2 $\leq \kt <$ 0.3 GeV}  &
{\bf 0.3 $\leq \kt <$ 0.4 GeV}  &
{\bf 0.4 $\leq \kt <$ 0.5 GeV}  &
{\bf 0.5 $\leq \kt <$ 0.6 GeV}
\cr
\hline
$N$                              &
$0.974 \pm 0.003 \pm 0.057$      &
$0.996 \pm 0.004 \pm 0.042$      &
$1.011 \pm 0.004 \pm 0.040$      &
$1.003 \pm 0.007 \pm 0.040$      &
$1.016 \pm 0.009 \pm 0.052$
\cr
$\lambda$                        &
$0.286 \pm 0.011 \pm 0.067$      &
$0.364 \pm 0.009 \pm 0.061$      &
$0.429 \pm 0.012 \pm 0.047$      &
$0.398 \pm 0.013 \pm 0.044$      &
$0.337 \pm 0.016 \pm 0.063$
\cr
$\Rto^2$ (fm$^2$)                &
$0.60 \pm 0.07 \pm 0.18$         &
$0.36 \pm 0.03 \pm 0.10$         &
$0.294 \pm 0.020 \pm 0.079$      &
$0.174 \pm 0.013 \pm 0.050$      &
$0.169 \pm 0.014 \pm 0.051$
\cr
$\Rts^2$ (fm$^2$)                &
$0.50 \pm 0.03 \pm 0.14$         &
$0.38 \pm 0.02 \pm 0.11$         &
$0.37 \pm 0.02 \pm 0.11$         &
$0.30 \pm 0.02 \pm 0.10$         &
$0.22 \pm 0.03 \pm 0.10$
\cr
$\Rlong^2$ (fm$^2$)              &
$1.09 \pm 0.11 \pm 0.37$         &
$0.72 \pm 0.04 \pm 0.17$         &
$0.82 \pm 0.05 \pm 0.16$         &
$0.60 \pm 0.04 \pm 0.15$         &
$0.75 \pm 0.06 \pm 0.18$
\cr
$\Rlongto^2$ (fm$^2$)            &
$-0.06 \pm 0.08 \pm 0.14$        &
$0.020 \pm 0.036 \pm 0.037$      &
$-0.065 \pm 0.028 \pm 0.031$     &
$-0.023 \pm 0.022 \pm 0.007$     &
$-0.121 \pm 0.024 \pm 0.065$
\cr
$\eto$ (GeV$^{-1}$)              &
$-0.091 \pm 0.004 \pm 0.060$     &
$-0.056 \pm 0.004 \pm 0.035$     &
$-0.037 \pm 0.004 \pm 0.027$     &
$-0.016 \pm 0.005 \pm 0.018$     &
$-0.003 \pm 0.006 \pm 0.018$
\cr
$\ets$ (GeV$^{-1}$)              &
$-0.123 \pm 0.004 \pm 0.071$     & 
$-0.130 \pm 0.004 \pm 0.061$     &
$-0.140 \pm 0.004 \pm 0.071$     &      
$-0.18 \pm 0.01 \pm 0.10$        & 
$-0.24 \pm 0.01 \pm 0.13$
\cr
$\elong$ (GeV$^{-1}$)            & 
$0.081 \pm 0.004 \pm 0.023$      & 
$0.048 \pm 0.004 \pm 0.017$      &
$0.019 \pm 0.005 \pm 0.015$      &      
$0.016 \pm 0.007 \pm 0.019$      & 
$-0.018 \pm 0.008 \pm 0.038$
\cr
\hline
$\chi^2$/DoF                     & 
$16389/15617$                    & 
$16080/15617$                    &
$15596/15617$                    & 
$15864/15617$                    & 
$15439/15617$
\cr
\hline
\hline
{\bf 0.8 $\leq |Y| <$ 1.6}       &
                                 &
                                 &
                                 &
                                 &
\cr
\hline
$N$                              & 
$0.972 \pm 0.003 \pm 0.049$      & 
$0.990 \pm 0.004 \pm 0.075$      &
$1.017 \pm 0.005 \pm 0.052$      & 
$1.019 \pm 0.007 \pm 0.057$      & 
$1.024 \pm 0.010 \pm 0.066$
\cr
$\lambda$                        & 
$0.315 \pm 0.008 \pm 0.070$      & 
$0.386 \pm 0.008 \pm 0.064$      &
$0.393 \pm 0.011 \pm 0.053$      & 
$0.379 \pm 0.013 \pm 0.055$      & 
$0.318 \pm 0.016 \pm 0.062$
\cr
$\Rto^2$ (fm$^2$)                & 
$0.62 \pm 0.04 \pm 0.20$         & 
$0.38 \pm 0.02 \pm 0.11$         &
$0.271 \pm 0.016 \pm 0.079$      & 
$0.204 \pm 0.014 \pm 0.062$      & 
$0.141 \pm 0.015 \pm0.053$
\cr
$\Rts^2$ (fm$^2$)                & 
$0.52 \pm 0.03 \pm 0.13$         & 
$0.38 \pm 0.02 \pm 0.11$         &
$0.34 \pm 0.02 \pm 0.10$         & 
$0.272 \pm 0.021 \pm 0.081$      & 
$0.226 \pm 0.026 \pm 0.079$
\cr
$\Rlong^2$ (fm$^2$)              & 
$1.06 \pm 0.08 \pm 0.35$         & 
$0.71 \pm 0.04 \pm 0.16$         &
$0.65 \pm 0.04 \pm 0.15$         & 
$0.64 \pm 0.05 \pm 0.16$         & 
$0.50 \pm 0.05 \pm 0.12$
\cr
$\Rlongto^2$ (fm$^2$)            & 
$0.019 \pm 0.055 \pm 0.076$      & 
$-0.029 \pm 0.026 \pm 0.036$     &
$-0.036 \pm 0.023 \pm 0.025$     & 
$-0.061 \pm 0.022 \pm 0.045$     & 
$-0.034 \pm 0.021 \pm 0.042$
\cr
$\eto$ (GeV$^{-1}$)              & 
$-0.070 \pm 0.004 \pm 0.049$     & 
$-0.046 \pm 0.004 \pm 0.035$     &
$-0.033 \pm 0.004 \pm 0.033$     & 
$-0.015 \pm 0.005 \pm 0.027$     & 
$0.007 \pm 0.007 \pm 0.009$
\cr
$\ets$ (GeV$^{-1}$)              & 
$-0.106 \pm 0.004 \pm 0.059$     & 
$-0.104 \pm 0.004 \pm 0.051$     &
$-0.131 \pm 0.005 \pm 0.064$     & 
$-0.161 \pm 0.005 \pm 0.088$     & 
$-0.23 \pm 0.01 \pm 0.14$
\cr
$\elong$ (GeV$^{-1}$)            & 
$0.066 \pm 0.004 \pm 0.026$      & 
$0.035 \pm 0.004 \pm 0.026$      &
$-0.003 \pm 0.005 \pm 0.036$     & 
$-0.028 \pm 0.006 \pm 0.047$     & 
$-0.060 \pm 0.009 \pm 0.072$
\cr
\hline
$\chi^2$/DoF                     & 
$15856/15617$                    & 
$15745/15617$                    &
$15658/15617$                    & 
$15895/15617$                    & 
$15592/15617$
\cr
\hline
\hline
{\bf 1.6 $\leq |Y| <$ 2.4}      &
                                &
                                &
                                &
                                &
\cr
\hline
$N$                             & 
$0.991 \pm 0.003 \pm 0.082$     & 
$1.019 \pm 0.005 \pm 0.069$     &
$1.066 \pm 0.005 \pm 0.078$     & 
$1.055 \pm 0.008\pm 0.074$      & 
$1.07 \pm 0.01 \pm 0.10$
\cr
$\lambda$                       & 
$0.261 \pm 0.008 \pm 0.079$     & 
$0.307 \pm 0.008 \pm 0.072$     &
$0.299 \pm 0.011 \pm 0.065$     & 
$0.264 \pm 0.014 \pm 0.074$     & 
$0.24 \pm 0.02 \pm 0.11$
\cr
$\Rto^2$ (fm$^2$)               & 
$0.54 \pm 0.04 \pm 0.19$        & 
$0.35 \pm 0.02 \pm 0.11$        &
$0.35 \pm 0.03 \pm 0.10$        & 
$0.219 \pm 0.017 \pm 0.064$     & 
$0.25 \pm 0.03 \pm 0.11$
\cr
$\Rts^2$ (fm$^2$)               & 
$0.53 \pm 0.03 \pm 0.15$        & 
$0.34 \pm 0.02 \pm 0.10$        &
$0.279 \pm 0.023 \pm 0.077$     & 
$0.229 \pm 0.026 \pm 0.072$     & 
$0.169 \pm 0.034 \pm 0.085$
\cr
$\Rlong^2$ (fm$^2$)             & 
$1.13 \pm 0.09 \pm 0.41$        & 
$0.61 \pm 0.04 \pm 0.17$        &
$0.78 \pm 0.06 \pm 0.19$        & 
$0.62 \pm 0.05 \pm 0.17$        & 
$0.54 \pm 0.07\pm 0.13$
\cr
$\Rlongto^2$ (fm$^2$)           & 
$-0.05 \pm 0.05 \pm 0.13$       & 
$0.012 \pm 0.029 \pm 0.033$     &
$-0.137 \pm 0.033 \pm 0.076$    & 
$-0.148 \pm 0.024 \pm 0.077$    & 
$-0.09 \pm 0.04 \pm 0.11$
\cr
$\eto$ (GeV$^{-1}$)             & 
$-0.102 \pm 0.004 \pm 0.070$    & 
$-0.063 \pm 0.004 \pm 0.048$    &
$-0.060 \pm 0.005 \pm 0.043$    & 
$-0.027 \pm 0.006 \pm 0.028$    & 
$0.16 \pm 0.01 \pm 0.12$
\cr
$\ets$ (GeV$^{-1}$)             & 
$-0.134 \pm 0.004 \pm 0.079$    & 
$-0.130 \pm 0.004 \pm 0.068$    &
$-0.167 \pm 0.005 \pm 0.082$    & 
$-0.19 \pm 0.01 \pm 0.11$       & 
$-0.31 \pm 0.01 \pm 0.17$
\cr
$\elong$ (GeV$^{-1}$)           & 
$0.045 \pm 0.004 \pm 0.056$     & 
$-0.003 \pm 0.005 \pm 0.045$    &
$-0.046 \pm 0.005 \pm 0.053$    & 
$-0.078 \pm 0.007 \pm 0.072$    & 
$-0.15 \pm 0.01 \pm 0.13$
\cr
\hline
$\chi^2$/DoF                    & 
$15966/15617$                   & 
$15866/15617$                   &
$15735/15617$                   & 
$15235/15617$                   & 
$15279/15617$
\cr
\hline
\end{tabular}
\caption{\sl{Results of the Bertsch-Pratt fits, Eq.~(7), to the
         experimental three-dimensional correlation functions
         $C'(\Ql,\Qts,\Qto)$
         over the range 0.04 $\leq \Ql,\Qts,\Qto \leq$ 1.0 GeV.
         The first errors are statistical and the second systematic.
         The quality of the fits is indicated
         by the value of $\chi^2$/DoF, which ranges from 0.98 to 1.05.}}
\label{bp}
\end{sidewaystable}

\newpage

\begin{sidewaystable}
\centering
\footnotesize
\begin{tabular}{|l||c|c|c|c|c|}
\hline
{\bf 0.0 $\leq |Y| <$ 0.8}       &
{\bf 0.1 $\leq \kt <$ 0.2 GeV}   &
{\bf 0.2 $\leq \kt <$ 0.3 GeV}   &
{\bf 0.3 $\leq \kt <$ 0.4 GeV}   &
{\bf 0.4 $\leq \kt <$ 0.5 GeV}   &
{\bf 0.5 $\leq \kt <$ 0.6 GeV}
\cr
\hline
$N$                              &
$0.993 \pm 0.003 \pm 0.010$      &
$1.004 \pm 0.003 \pm 0.008$      &
$0.985 \pm 0.004 \pm 0.017$      &
$0.946 \pm 0.005 \pm 0.041$      &
$0.85 \pm 0.01 \pm 0.14$
\cr
$\lambda$                        &
$0.266 \pm 0.012 \pm 0.067$      &
$0.358 \pm 0.009 \pm 0.056$      &
$0.441 \pm 0.012 \pm 0.035$      &
$0.440 \pm 0.013 \pm 0.023$      &
$0.482 \pm 0.017 \pm 0.069$
\cr
$v$                              &
$0.10 \pm 0.19 \pm 0.32$         &
$0.288 \pm 0.042 \pm 0.018$      &
$0.320 \pm 0.030 \pm 0.013$      &
$0.249 \pm 0.034 \pm 0.010$      &
$0.211 \pm 0.031 \pm 0.031$
\cr
$\Rz^2$ (fm$^2)$                 &
$-0.52 \pm 0.20 \pm 0.16$        &
$-0.184 \pm 0.044 \pm 0.075$     &
$-0.226 \pm 0.022 \pm 0.088$     &
$-0.203 \pm 0.013 \pm 0.081$     &
$-0.110 \pm 0.009 \pm 0.053$
\cr
$\Rt^2$ (fm$^2)$                 &
$0.75 \pm 0.04 \pm 0.17$         &
$0.50 \pm 0.02 \pm 0.12$         &
$0.41 \pm 0.02 \pm 0.10$         &
$0.313 \pm 0.011 \pm 0.081$      &
$0.193 \pm 0.007 \pm 0.048$
\cr
$\Rl^2$ (fm$^2)$                 &
$1.15 \pm 0.15 \pm 0.34$         &
$0.74 \pm 0.04 \pm 0.13$         &
$0.66 \pm 0.03 \pm 0.11$         &
$0.46 \pm 0.02 \pm 0.10$         &
$0.31 \pm 0.02 \pm 0.11$
\cr
$\dz$ (GeV$^{-1}$)               &
$-0.045 \pm 0.007 \pm 0.069$     &
$-0.014 \pm 0.006 \pm 0.052$     &
$0.008 \pm 0.007 \pm 0.019$      &
$0.061 \pm 0.008\pm 0.062$       &
$0.156 \pm 0.011 \pm 0.091$
\cr
$\dt$ (GeV$^{-1}$)               &
$-0.099 \pm 0.004 \pm 0.029$     &
$-0.089 \pm 0.005 \pm 0.023$     &
$-0.065 \pm 0.006 \pm 0.030$     &
$-0.066 \pm 0.008 \pm 0.036$     &
$-0.054 \pm 0.013 \pm 0.019$
\cr
$\dl$ (GeV$^{-1}$)               &
$0.038 \pm 0.002 \pm 0.098$      &
$0.014 \pm 0.002 \pm 0.090$      &
$0.005 \pm 0.003 \pm 0.051$      &
$-0.001 \pm 0.003 \pm 0.025$     &
$-0.002 \pm 0.005 \pm 0.012$
\cr
\hline
$\chi^2$/DoF                     &
$13583/11677$                    &
$16008/14375$                    &
$17555/16338$                    &
$18166/17554$                    &
$18702/18232$
\cr
\hline
\hline
{\bf 0.8 $\leq |Y| <$ 1.6}       &
                                 &
                                 &
                                 &
                                 &
\cr
\hline
$N$                              &
$0.948 \pm 0.005 \pm 0.021$      &
$0.977 \pm 0.006 \pm 0.006$      &
$0.964 \pm 0.009 \pm 0.010$      &
$0.914 \pm 0.012 \pm 0.051$      &
$0.859 \pm 0.022 \pm 0.072$
\cr
$\lambda$                        &
$0.324 \pm 0.008 \pm 0.065$      &
$0.380 \pm 0.010 \pm 0.057$      &
$0.425 \pm 0.013 \pm 0.035$      &
$0.464 \pm 0.019 \pm 0.023$      &
$0.464 \pm 0.034 \pm 0.028$
\cr
$v$                              &
$0.754 \pm 0.022 \pm 0.061$      &
$0.782 \pm 0.014 \pm 0.036$      &
$0.742 \pm 0.017 \pm 0.027$      &
$0.777 \pm 0.014 \pm 0.031$      &
$0.743 \pm 0.023 \pm 0.040$
\cr
$\Rz^2$ (fm$^2)$                 &
$-0.187 \pm 0.054 \pm 0.076$     &
$-0.104 \pm 0.024 \pm 0.051$     &
$-0.114 \pm 0.016 \pm 0.054$     &
$-0.102 \pm 0.013 \pm 0.051$     &
$-0.106 \pm 0.011 \pm 0.052$
\cr
$\Rt^2$ (fm$^2)$                 &
$0.56 \pm 0.02 \pm 0.15$         &
$0.39 \pm 0.01 \pm 0.11$         &
$0.32 \pm 0.01 \pm 0.10$         &
$0.235 \pm 0.010 \pm 0.071$      &
$0.164 \pm 0.009 \pm 0.049$
\cr
$\Rl^2$ (fm$^2)$                 & 
$0.83 \pm 0.06 \pm 0.27$         &
$0.58 \pm 0.03 \pm 0.13$         &
$0.46 \pm 0.03 \pm 0.10$         &
$0.43 \pm 0.03 \pm 0.12$         &
$0.294 \pm 0.032 \pm 0.085$
\cr
$\dz$ (GeV$^{-1}$)               &
$-0.07 \pm 0.01 \pm 0.10$        &
$0.00 \pm 0.01 \pm 0.12$         &
$0.071 \pm 0.010 \pm 0.053$      &
$0.106 \pm 0.012 \pm 0.076$      &
$0.125 \pm 0.019 \pm 0.084$
\cr
$\dt$ (GeV$^{-1}$)               &
$-0.068 \pm 0.007 \pm 0.041$     &
$-0.075 \pm 0.008 \pm 0.033$     &
$-0.069 \pm 0.011 \pm 0.043$     &
$-0.046 \pm 0.015 \pm 0.036$     &
$-0.079 \pm 0.027 \pm 0.079$
\cr
$\dl$ (GeV$^{-1}$)               &
$0.099 \pm 0.010 \pm 0.078$      &
$0.022 \pm 0.008 \pm 0.099$      &
$-0.032 \pm 0.008 \pm 0.062$     &
$-0.058 \pm 0.010 \pm 0.047$     &
$-0.077 \pm 0.014 \pm 0.060$
\cr
\hline
$\chi^2$/DoF                     &
$8624/7139$                      &
$9778/8788$                      &
$11004/9870$                     &
$11365/10518$                    &
$11603/10885$
\cr
\hline
\hline
{\bf 1.6 $\leq |Y| <$ 2.4}       &
                                 &
                                 &
                                 &
                                 &
\cr
\hline
$N$                              &
$0.899 \pm 0.016 \pm 0.054$      &
$0.963 \pm 0.020 \pm 0.064$      &
$0.902 \pm 0.021 \pm 0.040$      &
$0.888 \pm 0.028 \pm 0.052$      &
$0.48 \pm 0.01 \pm 0.39$
\cr
$\lambda$                        &
$0.342 \pm 0.019 \pm 0.078$      &
$0.354 \pm 0.022 \pm 0.072$      &   
$0.454 \pm 0.012 \pm 0.041$      &
$0.438 \pm 0.038 \pm 0.030$      &
$1.26 \pm 0.04 \pm 0.63$
\cr
$v$                              &
$0.893 \pm 0.012 \pm 0.044$      &
$0.931 \pm 0.008 \pm 0.047$      &
$0.927 \pm 0.009 \pm 0.033$      &
$0.912 \pm 0.012 \pm 0.042$      &
$0.93 \pm 0.04 \pm 0.11$
\cr
$\Rz^2$ (fm$^2)$                 &
$0.031 \pm 0.043 \pm 0.029$      &
$0.006 \pm 0.032 \pm 0.022$      &
$0.015 \pm 0.026 \pm 0.025$      &
$0.020 \pm 0.030 \pm 0.026$      &
$-0.034 \pm 0.024 \pm 0.019$
\cr
$\Rt^2$ (fm$^2)$                 &
$0.39 \pm 0.02 \pm 0.12$         &
$0.291 \pm 0.017 \pm 0.087$      &
$0.172 \pm 0.011 \pm 0.052$      &
$0.159 \pm 0.014 \pm 0.048$      &
$0.071 \pm 0.005 \pm 0.021$
\cr
$\Rl^2$ (fm$^2)$                 &
$0.42 \pm 0.05 \pm 0.16$         &
$0.42 \pm 0.05 \pm 0.15$         &
$0.34 \pm 0.03 \pm 0.10$         &
$0.38 \pm 0.06 \pm 0.10$         &
$0.12 \pm 0.03 \pm 0.11$
\cr
$\dz$ (GeV$^{-1}$)               &
$0.03 \pm 0.04 \pm 0.12$         &
$0.060 \pm 0.029 \pm 0.080$      &
$-0.014 \pm 0.033 \pm 0.021$     &
$0.196 \pm 0.031 \pm 0.075$      &
$0.00 \pm 0.03 \pm 0.16$
\cr
$\dt$ (GeV$^{-1}$)               &
$-0.033 \pm 0.023 \pm 0.082$     &
$-0.087 \pm 0.024 \pm 0.093$     &
$-0.030 \pm 0.009 \pm 0.050$     &
$-0.070 \pm 0.032 \pm 0.055$     &
$0.216 \pm 0.031 \pm 0.071$
\cr
$\dl$ (GeV$^{-1}$)               &
$-0.001 \pm 0.036 \pm 0.066$     &
$-0.049 \pm 0.028 \pm 0.081$     &
$0.025 \pm 0.030 \pm 0.054$      &
$-0.162 \pm 0.029 \pm 0.065$     &
$0.02 \pm 0.03 \pm 0.16$
\cr
\hline
$\chi^2$/DoF                     &
$4168/3804$                      &
$5110/4648$                      &
$5952/5169$                      &
$5775/5490$                      &
$5876/5642$
\cr
\hline
\end{tabular}
\caption{\sl{Results of the Yano-Koonin fits, Eq.~(8), to the
         experimental three-dimensional correlation functions
         $C'(\qt,\ql,\qz)$
         over the range 0.04 $\leq \qt,\ql,\qz \leq$ 1.0 GeV.
         The first errors are statistical and the second systematic.
         The quality of the fits is indicated
         by the value of $\chi^2$/DoF, which ranges from 1.03 to 1.20.}}
\label{yk}
\end{sidewaystable}

\newpage

\begin{sidewaystable}
\centering
\normalsize
\begin{tabular}{|c||c|c||c|c|}
\hline
 & \multicolumn{2}{c||}{} & \multicolumn{2}{c|}{} \\
 & \multicolumn{2}{c||}{\bf \Large{\bf BP radii}} & 
 \multicolumn{2}{c|}{\bf \Large{\bf YK radii}} \\
 & \multicolumn{2}{c||}{} & \multicolumn{2}{c|}{} \\
                               &
{\bf $d \Rlong^2 / d \kt$}     &
{\bf $d \Rts^2 / d \kt$}       &
{\bf $d \Rl^2 /d \kt$}         &
{\bf $d \Rt^2 /d \kt$}
\cr
                            &
\footnotesize{(fm$^2$/GeV)} &
\footnotesize{(fm$^2$/GeV)} &
\footnotesize{(fm$^2$/GeV)} &
\footnotesize{(fm$^2$/GeV)}
\cr
 & & & &
\cr
\hline
\hline
 & & & &
\cr
{\bf $|Y| < $ 0.8} & $-0.46 \pm 0.20 \pm 0.35$ & $-0.59 \pm 0.08 \pm 0.19$ &
$-1.60 \pm 0.13 \pm 0.38$ & $-1.14 \pm 0.05 \pm 0.23$
\cr
 & & & &
\cr
\hline
 & & & &
\cr
{\bf 0.8 $\leq |Y| <$ 1.6} & $-0.91 \pm 0.18 \pm 0.30$ &
$-0.66 \pm 0.08 \pm 0.15$ &
$-1.04 \pm 0.12 \pm 0.23$ & $-0.84 \pm 0.04 \pm 0.15$
\cr
 & & & &
\cr
\hline
 & & & &
\cr
{\bf 1.6 $\leq |Y| <$ 2.4} & $-0.64 \pm 0.21 \pm 0.36$ & 
$-0.80 \pm 0.09 \pm 0.28$ &
$-0.82 \pm 0.13 \pm 0.17$ & $-0.70 \pm 0.04 \pm 0.20$
\cr
 & & & &
\cr
\hline
\end{tabular}
\caption{\sl{Slopes of the linear fits to the dependence of the longitudinal
             and transverse squared radii of the BP and YK parameterizations
             on $\kt$.
             Input to the fits are the measured values of
             $\Rlong^2$, $\Rts^2$, $\Rl^2$ and $\Rt^2$,
             reported in Tables~\ref{bp} and~\ref{yk}.
             The first errors are statistical and the second
             systematic.}}
\label{slope}
\end{sidewaystable}

\end{document}